\documentclass[useAMS,usenatbib]{mn2e}

\usepackage{graphicx}
\title[Spectral properties of the narrow-line region in Seyfert galaxies
selected from the SDSS-DR7]{Spectral properties of the narrow-line region in Seyfert galaxies
selected from the SDSS-DR7}
\author[L. Vaona, et al.]{L. Vaona, S. Ciroi\thanks{E-mail:
stefano.ciroi@unipd.it}, F. Di Mille, V. Cracco, G. La Mura, P. Rafanelli\\
Department of Astronomy, Padova University, vicolo dell'Osservatorio 3, I-35122 Padova, Italy}

\begin{document}

\date{Accepted ****. Received ****; in original form ****}

\pagerange{\pageref{firstpage}--\pageref{lastpage}} \pubyear{****}

\maketitle

\label{firstpage}

\begin{abstract}
Although the properties of the narrow-line region (NLR) of active galactic nuclei (AGN) have been deeply studied by many authors in the past three decades, many questions are still open. The main goal of this work is to explore the NLR of Seyfert galaxies by collecting a large statistical spectroscopic sample of Seyfert 2 and Intermediate-type Seyfert galaxies having a high signal-to-noise ratio in order to take advantage of a high number of emission-lines to be accurately measured.

2153 Seyfert 2 and 521 Intermediate-type Seyfert spectra were selected from Sloan Digital Sky Survey - Data Release 7 (SDSS-DR7) with a 
diagnostic diagram based on the oxygen emission-line ratios. All the emission-lines, broad components included, were measured by means of a self-developed code, after the subtraction of the stellar component. Physical parameters, such as internal reddening, ionization parameter, temperature, density, gas and stellar velocity dispersion were determined for each object. Furthermore, we estimated mass and radius of the NLR, kinetic energy of the ionized gas, and black-hole accretion rate. 

From the emission-line analysis and the estimated physical properties, it appears that the NLR is similar in Seyfert 2 and Intermediate-Seyfert galaxies. The only differences, lower extinction, gas kinematics in general not dominated by the host galaxy gravitational potential and higher percentage of [O\,{\sc iii}]$\lambda$5007 blue asymmetries in Intermediate-Seyfert can be ascribed to an effect of inclination of our line of sight with respect to the torus axis. 
\end{abstract}


\section{Introduction}

Seyfert galaxies are characterized by a bright star-like nucleus, whose spectrum shows emission-lines covering a wide range of ionization stages.  
The main components of a Seyfert nucleus are: a central source, the broad-line region (BLR), a molecular dusty torus, the narrow-line Region (NLR) and a possible extended NLR (ENLR). The BLR has a typical sub-parsec size and it is surrounded by an optically thick torus of dust. 
The NLR is characterized by low density clouds ($N_{\rm e}\sim$ 10$^{2}$--10$^{6}$ cm$^{-3}$), temperature $T_{\rm e}\sim10^4$ K, size $\rm \sim 1~kpc$ and full-width at half maximum (FWHM) of the emission-lines in the range $\sim$ 200--1000 km s$^{-1}$. 
The torus confines the escaping ionizing radiation into two oppositely directed cones \citep[see e.g.][]{sb92,kriss97,pogge93,capetti99,tsve92,ferruit99,miy92,vei93,falc96,wil94}. 
Inside the cones, the gas is directly ionized by a non-thermal power-law spectrum from the central source, even if the presence of shocks \citep{dop95} and photoionization by star-forming regions close to the nucleus \citep[composite galaxies,][]{hill99} cannot be excluded. 
This model, called Unified Model and based on the spectropolarimetric analysis of NGC 1068, performed by \citet{ant85}, is the most accepted active galactic nuclei (AGN) picture \citep{ant93}. The Unified Model is able to explain the main different characteristics of Seyfert 1 and Seyfert 2 galaxies. In the first case, the torus is seen face on, then the BLR is directly observable, in the second case, the line of sight intercepts the torus and the BLR is not visible. 
Deep CCD images of few nearby galaxies made through narrow band interference filters revealed an extended structure very rich of gas clouds surrounding the active nucleus \citep{pogge88,pogge89,tad89}.
According to \citet{1996ApJ...463..498S} Seyfert 1 galaxies seem to have NLRs with a much smaller size than those of Seyfert 2 if observed along the torus axis. But \citet{1998ApJ...499..719K} stressed that this cannot be explained by a simple orientation effect and that NLRs of Seyfert 1 are physically compact. However, \citet{2003ApJ...597..768S} showed that both Seyfert types have similar distributions of NLR sizes, and elongated shapes are observed more frequently in Seyfert 2 than in Seyfert 1, whose emission is more concentrated toward the nucleus. These results are in agreement with the Unified Model if we assume that the NLR gas has a disk-like distribution.
With the advent of the {\it Hubble Space Telescope (HST)}, it was possible to observe that ionization cones \citep[][and references therein]{afa07_a, wilson97} are in many cases closely associated with radio jets and lobes \citep{wil94}, and dust lanes \citep{martini03a,martini03b}.

If we accept the Unified Model, we should expect that the spectroscopic properties of the NLR are similar in Seyfert 2, intermediate-type Seyfert and Seyfert 1 galaxies, with an increasing contribution from highly ionized gas closer to the central source. Unfortunately, even in case of high signal-to-noise spectra with good resolution, the narrow components of the recombination emission lines in Seyfert 1 galaxies are difficult to measure, since they are included in and hidden
by the broad and bright components emitted by the BLR. This is even harder when the application of automatic fitting procedures is mandatory because of the analysis of large samples. Spectroscopic differences and similarities among Seyfert types have been found in visible spectra, but often limited to single objects 
\citep[see e.g.][]{1998ApJ...499..719K, 2000ApJ...532..867F}, or small samples \citep[see e.g.][]{bennert06a, bennert06b}, while large samples are very rarely studied \citep{zhang08}.

The aim of this work is to explore possible differences between Seyfert 2 and Intermediate-type Seyfert galaxies, through a detailed analysis of a large sample of Seyfert spectra and taking advantage also of the weak lines, which are too often neglected.
This work is organized in five sections: in Section 2 the selection criteria of the spectra are described; Section 3 describes the measurements of the spectra and their classification in Seyfert 2 and Intermediate-type Seyfert. The emission-line analysis and the NLR physical characteristics are dealt in Section 4 and 5 respectively. The last section summarizes the main achievements of this work.

\section{The spectra selection}
To build a statistical sample of Seyfert galaxies, we decided to exploit the Sloan Digital Sky Survey - Data Release 7 (SDSS-DR7) database \citep{dr7_ref}. 
We used a diagnostic diagram (hereafter $\rm O_{123}$) based on the [O\,{\sc ii}]$\lambda$3727/[O\,{\sc iii}]$\lambda$5007 (hereafter $\rm O_{23}$) and [O\,{\sc i}]$\lambda$6300/[O\,{\sc iii}]$\lambda$5007 (hereafter $\rm O_{13}$) ratios. The first ratio gives a measure of the ionization level, while the second ratio is a tracer of the hardness of the ionizing spectrum at large radii from the source \citep{sf90}. 
Up to now this ratio has never been employed to classify a large number of objects. 
The great advantages of the $\rm O_{123}$ diagram are the following: i) the oxygen lines are visible in all AGN spectra, ii) oxygen is the only element for which three different states of ionization exist in the visible range, iii) the employed ratios are sensitive to the ionization parameter and power-law index and are weakly dependent on the metallicity and density, iv) they are not contaminated by stellar spectral features, therefore it is not necessary to subtract the stellar template from the original spectra before measuring the lines, v) the lines are not contaminated by traditional BLR, vi) and finally [O\,{\sc i}]$\lambda$6300 (and to a lesser extent [N\,{\sc ii}]$\lambda$6548,6584) are more sensitive to the hard AGN continuum compared with the stellar continuum (of H{\sc ii} regions). 
Nevertheless, this diagram has two main disadvantages: the ratios are not reddening free and the [O\,{\sc i}]$\lambda$6300 is sometimes weak and difficult to be measured especially when the signal-to-noise ratio (S/N) of the [O\,{\sc i}]$\lambda$6300 is less than 10.
From the SDSS-DR7 we extracted all galaxies showing [O\,{\sc ii}]$\lambda$3727, [O\,{\sc iii}]$\lambda$5007, [O\,{\sc i}]$\lambda$6300 emission-lines, but with the constraint S/N([O\,{\sc i}]$\lambda$6300)$>3$, obtaining 119226 targets. 

\begin{figure}
  \centering
 \includegraphics[width=84 mm]{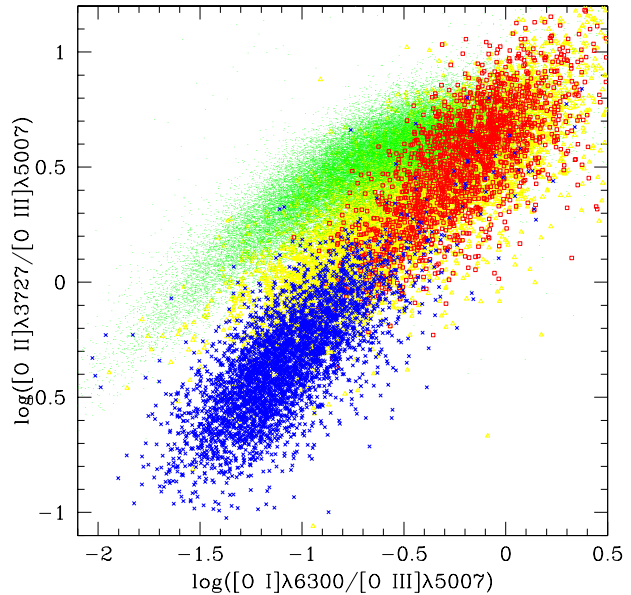}
  \caption{\small Kewley classification \citep{k06} on the $\rm O_{123}$ diagram applied to the sample of 62036 objects: blue crosses are Seyfert 2, red squares are LINERs, green dots are star-forming galaxies and yellow triangles are composite galaxies.}
 \label{fig:O123_class}
\end{figure}

To check the effectiveness of oxygen spectral lines in isolating Seyfert galaxies, we carried out a test. First, we exploited the database of $85224$ galaxies published by \citet{k06}, finding 62036 galaxies in common with our sample. 
Then, we plotted the Veilleux-Osterbrock diagnostic diagrams \citep[hereafter VO]{vo87} and we used the relations given in \citet{k06} to classify these objects as Seyfert, Low-Ionization Nuclear Emission-line Regions (LINERs), star-forming and composite Seyfert-H\,{\sc ii} galaxies. Since the \citet{k06} data are reddening corrected and the host galaxy spectrum is removed, they are not directly comparable with our sample. Therefore, we decided to extract from SDSS the spectroscopic information about these already classified objects and we plotted the $\rm O_{123}$ diagram (Fig.\,\ref{fig:O123_class}). The result was extremely encouraging: this plot shows a V-shape where the upper side is populated by star-forming galaxies, while the lower side is populated by Seyfert galaxies and LINERs, which are sufficiently separated. This diagram is not new, since it was already published by \citet{sf90} (see their fig.\,13), but the areas indicated by these authors, where Seyfert galaxies, LINERs and H\,{\sc ii} regions are expected to be found, do not perfectly match our data. 

\begin{figure}
  \centering
 \includegraphics[width=84 mm]{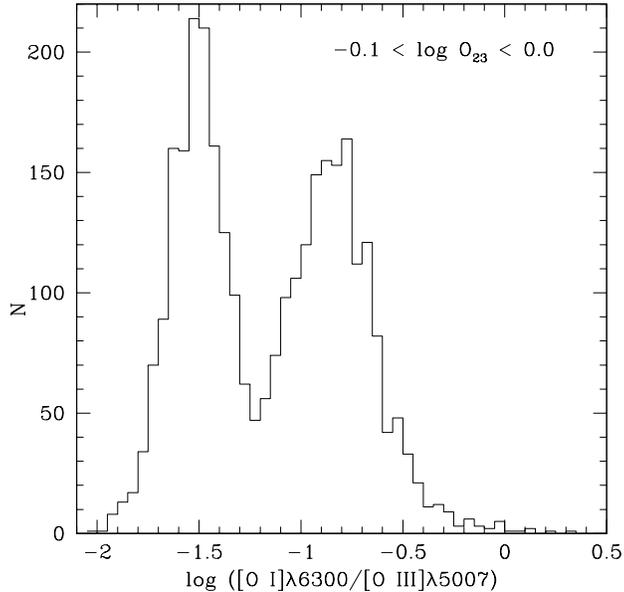}
  \caption{\small Histogram of $\log(\rm O_{13})$ values with $\log(\rm O_{23})$ in the range [$-0.1$,$0$].}
 \label{fig:O123_strip}
\end{figure}

From this diagram we derived an empirical separation line to isolate Seyfert galaxies. We sampled $\rm \log(O_{23})$ in bins of $0.1$ dex, in the range $[-1.2, 0.3]$ and plotting histograms of $\rm \log(O_{13})$ in bins of $0.1$ dex, we estimated the value of the minimum between the two peaks of the distributions for each bin (as shown in Fig.\,\ref{fig:O123_strip}). Finally, taking into account the mean point of the bin,  we interpolated these data with a polynomial function, obtaining as result the following formula:
 \begin{equation}
\rm 
\log(O_{23})=0.20-0.25\times \log(O_{13})-0.39\times [\log(O_{13})]^{2}.
\label{eq1}
\end{equation}
\begin{figure}
  \centering
 \includegraphics[width=84 mm]{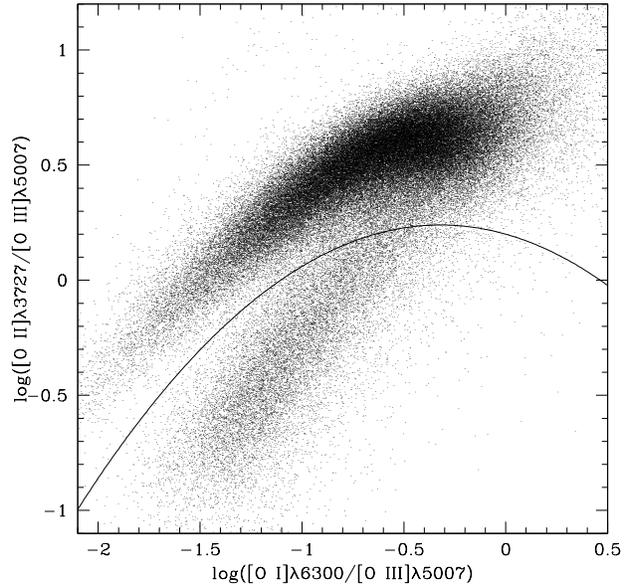}
  \caption{\small The O$_{123}$ diagram of our sample of 119226 objects. The solid line is the empirical separation described in Eq.\,\ref{eq1}.}
\label{fig:O123_table134}
\end{figure}
By selecting the objects under the curve we include 97 per cent of the original sample of Seyfert galaxies. Of course this new sub-sample is contaminated by other objects: in particular, it contains 60 per cent of Seyfert galaxies and 31 per cent of composite Seyfert-H\,{\sc ii} galaxies, but only 4 per cent of LINERs and 5 per cent of star-forming galaxies. So, we can conclude that about 10 per cent of a sample selected by following this method consists of non-Seyfert galaxies.
It is interesting to note that the lower sequence contains both narrow and broad-lined AGN. This is due to the fact that it uses only the oxygen lines. In conclusion the $\rm O_{123}$ diagram is useful for the classification of emission-line galaxies in general and not only for narrow emission-line ones. 
By applying Eq.\,\ref{eq1} to the whole sample of 119226 objects (Fig.\,\ref{fig:O123_table134}) we found that about 16000 populate the Seyfert region on the $\rm O_{123}$ diagram. 
Since the SDSS spectra are obtained with a $3$ arcsec fibre aperture, a redshift limit $z\leq0.1$ was adopted to reduce the flux contamination by extra-nuclear sources \citep{k06}. The lower limit $z\geq0.02$ is required because the [O\,{\sc ii}]$\lambda$3727 line must be visible in the 3800--9200 \AA\ spectral range covered by the SDSS detectors.

\begin{figure}
  \centering
 \includegraphics[width=84 mm]{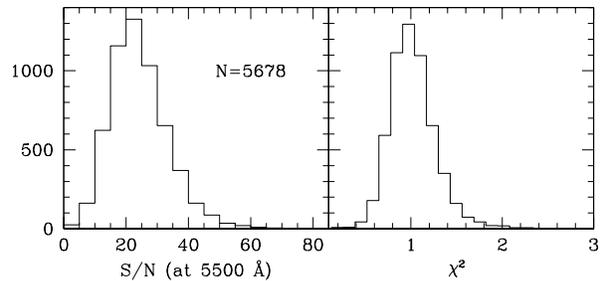}
 \caption{\small $\rm S/N$ ratio of the continuum at rest frame 5500 \AA\ for the $5678$ spectra of our sample (left). \small $\rm \chi^2/dof$ distribution of {\sc starlight} fittings to the 5678 spectra (right).}
 \label{fig:sn5500}
\end{figure}

The total number of galaxies in the Seyfert region that satisfy the mentioned conditions is 5678. 
The left panel of Fig.\,\ref{fig:sn5500} shows the S/N ratio of the spectra measured at rest-frame 5500 \AA. In most cases (91 per cent) it is between $10$ and $40$, with a peak around $20$.

\section{Measurements and spectra classification}\label{sec:spectra}
The VO diagrams are sensitive to the measurement of the H$\beta$ emission-line that is made difficult by the presence of the underlying stellar absorption. A suitable correction is mandatory to avoid an over-estimate of the [O\,{\sc iii}]$\lambda$5007/H$\beta$ diagnostic ratio. 
The fitting and subtraction of the underlying stellar continuum from each spectrum was carried out with the spectral synthesis code {\sc starlight} \citep{cid05,cid07}. This code makes a linear combination of synthetic spectra which are reddened and then convolved with a broadening function, allowing to obtain the stellar velocity dispersion. 

\begin{figure*}
  \centering
\includegraphics[width=168 mm]{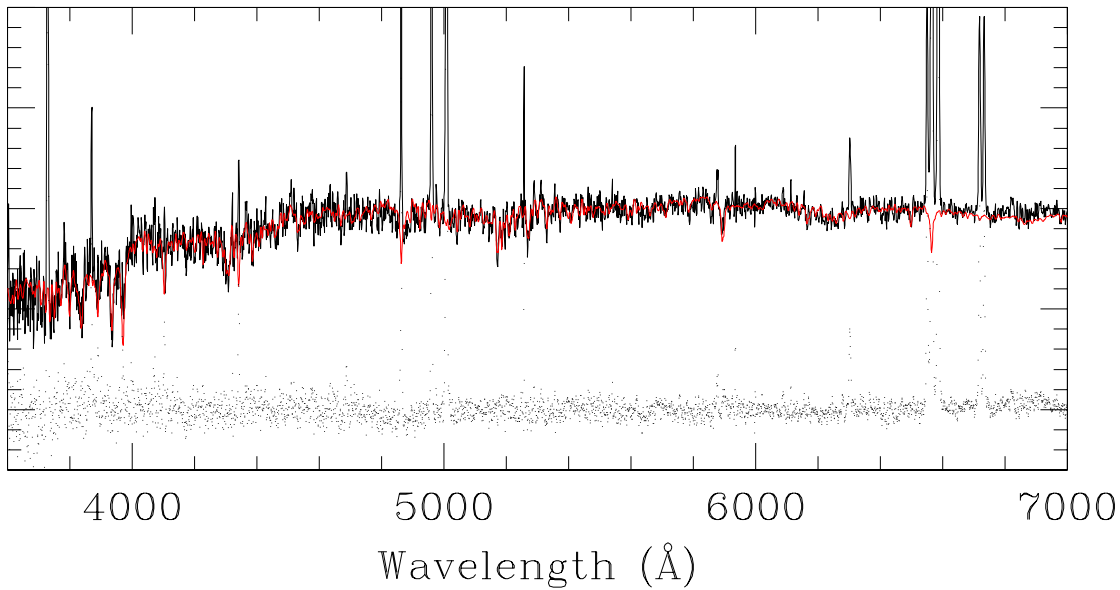}
  \caption{\small An example of correction for stellar component. The original spectrum is plotted in black, the fit to the stellar component in red and the residual pure emission-line spectrum in black, dotted line.}
 \label{fig:starlightfit_tot}
\end{figure*}

Before being analysed with {\sc starlight}, the spectra were processed with {\sc iraf}. A correction for Galactic absorption was first applied to the spectra, by using {\sc deredden} and A$(V)$ extinction values given by NASA/IPAC Extragalactic Database (NED). Then, the spectra were shifted to rest-frame with {\sc newredshift} by using z values given by SDSS, re-gridded to a dispersion of $1$ \AA/px with {\sc dispcor} and converted into text format.
We used as base $92$ synthetic spectra from \citet{bru03} by combining $23$ ages (from $\rm 10^{6}\,yr$  up to 
$\rm 13 \times 10^{9}\,yr$) with $4$ metallicities ($Z=0.004, 0.008, 0.02$ and $0.05$), and \citet{ccm89} (hereafter CCM) as extinction function. We masked the emission-lines in order to improve the quality of the fit. 
The goodness of the {\sc starlight} fitting evaluated with $\chi^{2}$ is peaked around $1$ (Fig.\,\ref{fig:sn5500}, right panel). The best-fitting synthetic spectrum of the continuum for each galaxy was subtracted from the observed one in order to obtain a pure emission-line spectrum, where hydrogen and helium Balmer lines could be correctly measured (Fig.\,\ref{fig:starlightfit_tot}). 

The emission-lines were fitted by means of a dedicated code written in C language. This code, named {\sc ggfit}, was developed within our group, in order to fit in automatic way the emission-lines of a group of spectra with a multi-Gaussian non-linear least-squares fitting method.
In the case of the blended lines [N\,{\sc ii}]$\lambda$6548,6584 and H$\alpha$ or [S\,{\sc ii}]$\lambda$6716,6731, the code fits these features simultaneously. 
The output is a table collecting 12 parameters for each spectral line:
\begin{itemize}
 \item the central wavelength ($x_{0}$), amplitude ($A$), sigma ($\sigma$), and their errors ($\Delta x_{0}$, $\Delta A$,  $\Delta \sigma$); 
 \item the measured flux ($F_{\rm M}$);
 \item the $\rm S/N$ ratio;
 \item the R$^{2}$ parameter, the sum of the square residuals, defined as R$^{2}=(\delta x_{0})^2+(\delta A)^2+(\delta \sigma)^2$, where $\delta x_{0}$, $\delta A$ and $\delta \sigma$ are the differences between the last iteration values and the previous ones;
 \item the Gaussian flux ($F_{\rm G}$) and its error ($\Delta F_{\rm G}$);
 \item the $\chi^{2}$ of the fitting.
\end{itemize}
The first step is to determine the line boundaries, i. e. the wavelengths at which the line profile fades into the continuum on both sides of the line. For this purpose, the algorithm evaluates the intensity of the line profile ($I(\lambda)$) from the peak towards bluer wavelengths until it reaches a minimum ($I_{\rm min}$). If $I_{\rm min} < I_{\rm c} + \sigma_{\rm c}$, where $I_{\rm c}$ and $\sigma_{\rm c}$ are the average continuum intensity and its error, 
the associated wavelength is assumed as the blue boundary of the line, otherwise the algorithm searches for a new minimum. The same procedure is repeated on the red side of the line. 
In order to prevent non-convergence, for each line or blend of lines a maximum value of the baseline is imposed. Each interval was empirically chosen to avoid nearby lines or features and the regions where the continuum was evaluated.    
If this value is reached, the algorithm starts again the procedure by comparing $I(\lambda)$ with $I_{\rm c} + 2\sigma_{\rm c}$. If the convergence is not reached, then it tries with $3\sigma_{\rm c}$. 

\begin{figure}
 \includegraphics[width=84 mm]{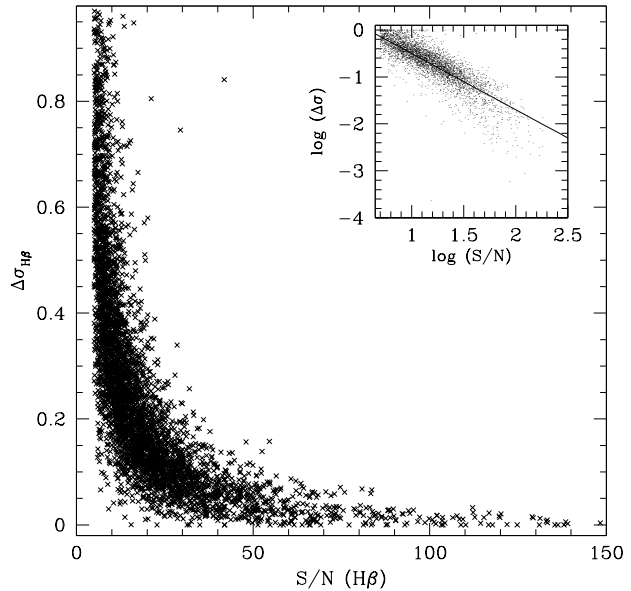}
 \caption{\small $\rm S/N (H\beta)$ vs. $\rm \Delta\sigma (H\beta)$ for the measured $5678$ spectra. The same in the top-right panel, but in logarithmic scale.}
 \label{fig:log_errsig_Hb}
\end{figure}

Once the boundaries are determined, the flux is obtained by integrating the spectral feature. 
The flux error is calculated by means of $\Delta A$ and $\Delta \sigma$. $\Delta A$ is the value of $\sigma_{\rm c}$ determined in proximity of the emission-line, while to estimate $\Delta \sigma$ a Monte Carlo method is applied during the fitting procedure.
In particular, starting from the Gaussian parameters, the code generates 3000 possible models near the best solution, within the fixed intervals: $A-\sigma_{\rm c}<A<A+\sigma_{\rm c}$, $\sigma-1<\sigma<\sigma+1$, $x_{0}-1<x_{0}<x_{0}+1$, and changing randomly the values of the parameters. 
A $\chi^{2}$ test is performed and only the models with a significance level above 70 per cent are considered. 
This procedure was applied only to features with one or two lines, for computational time reasons. 
$\Delta A$ is always lower than $\sigma_{\rm c}$, whereas $\Delta \sigma$ and $\Delta x_0$ both depend on the $\rm S/N$. 
In Fig.\,\ref{fig:log_errsig_Hb} the case of H$\beta$ is presented. 
As it can be seen, $\Delta \sigma \propto \rm(S/N)^{\alpha}$. Following \citet{cor99} and by applying a linear fitting to $\log \Delta \sigma$ vs. $\log \rm S/N$, a simple relation to estimate $\Delta \sigma$ can be derived. In this case, $\Delta \sigma$ is given by:
\begin{equation}
\log(\Delta\sigma)=-1.2 \log({\rm S/N})+0.7
\label{equ:errsig}
\end{equation}
but similar results are obtained for all the measured lines. 
$\Delta x_{0}$ is similar to $\Delta\sigma$ and it correlates well with $\Delta\sigma$ and $\rm S/N$. The correlation coefficient of the interpolation is $0.95$. The typical wavelength error is about $0.5$ \AA\ when the $\rm S/N$ is around $10$.

\begin{figure}
\includegraphics[width=84 mm]{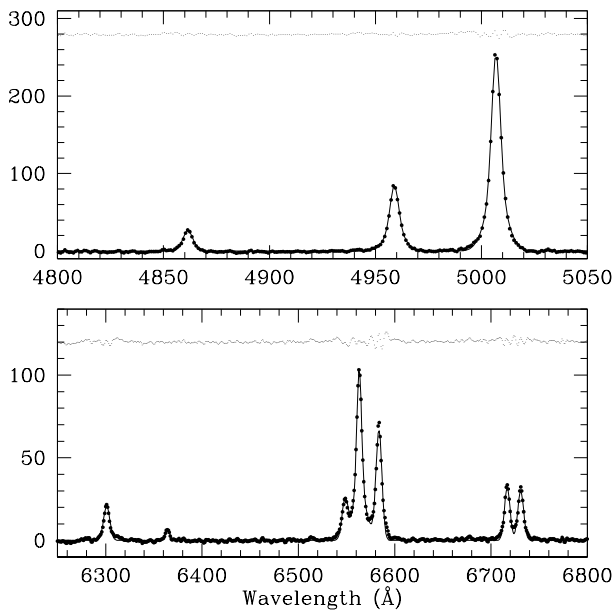}
\caption{\small Examples of lines fitting. The observed spectrum (black dots) is overlapped to the results of the fitting (solid line). Residuals are plotted on the top of the diagrams (dotted line).}
\label{fig:linefit}
\end{figure}

In order to check the reliability of the automatic fitting procedure, about $1300$ spectra were measured by hand. 
The line boundaries limits and the fluxes obtained are the same with very rare exceptions. The number of failures increases when the $\rm S/N<5$, therefore only the lines with a $\rm S/N>5$ were measured. They are listed in Table \ref{tab:elenco_comp_a}, with the maximum number of components adopted in the fitting, the ionization potentials and the logarithm of the critical density \citep{derob86,mendoza83}. No other line was taken into consideration because it was too weak to be detected, as [Fe\,{\sc x}]$\lambda$6374, $\rm H\epsilon$ and higher order Balmer lines, or in blend but presumably giving a small contribution to the total flux of the blended line, as [S\,{\sc iii}]$\lambda$6312, [Ca\,{\sc v}]$\lambda$6087 \citep{nagao03}.
Our software was able to fit successfully 4254 spectra: the others were rejected because either the most important emission-lines (i.e. those used in diagnostic diagrams) had $\rm S/N < 5$ or (in a few cases) their fitting did not converge. Fig.\,\ref{fig:linefit} shows an example of the fitting.

\begin {table}
\caption{Measured lines.}
\begin{center}
 {\small
 \begin{tabular}{|llccc|}
 \hline 
   line& comp & IP low eV & IP high eV & $\log(N_{\rm c} \,{\rm cm^{-3}})$\\
 \hline
\ [O\,{\sc ii}]$\lambda$3727     &   1&  13.62   &     35.12     &     3.7\\
\ [Ne\,{\sc iii}]$\lambda$3869   &   1&  40.96   &     63.45     &     7.0\\
\ [Ne\,{\sc iii}]$\lambda$3967   &   1&  40.96   &     63.45     &     7.0\\
\ [S\,{\sc ii}]$\lambda$4070     &   1&  10.36   &     23.34     &     6.4\\
\ H$\delta$\,$\lambda$4102       &   1&  13.60   &      --       &      --\\
\ H$\gamma$\,$\lambda$4340       &   1&  13.60   &      --       &      --\\
\ [O\,{\sc iii}]$\lambda$4363    &   1&  35.12   &     54.94     &     7.5\\
\ He\,{\sc ii}\,$\lambda$4686    &   1&  24.59   &     54.42     &      --\\
\ [Ar\,{\sc iv}]$\lambda$4711    &   1&  40.74   &     59.81     &     4.4\\
\ [Ar\,{\sc iv}]$\lambda$4740    &   1&  40.74   &     59.81     &     5.6\\
\ H$\beta$\,$\lambda$4861        &   2&  13.60   &      --       &      --\\
\ [O\,{\sc iii}]4959    &   2&  35.12   &     54.94     &     5.8\\
\ [O\,{\sc iii}]$\lambda$5007    &   2&  35.12   &     54.94     &     5.8\\
\ [N\,{\sc i}]$\lambda$5199      &   1&  0       &     14.53     &     3.3\\
\ [Fe\,{\sc vii}]$\lambda$5721   &   1&  99.1    &    124.98     &     7.6\\
\ He\,{\sc i}\,$\lambda$5876     &   1&  0       &     24.59     &      --\\ 
\ [Fe\,{\sc vii}]$\lambda$6087   &   1&  99.1    &    124.98     &     7.6\\
\ [O\,{\sc i}]$\lambda$6300      &   1&  0       &     13.62     &     6.3\\ 
\ [O\,{\sc i}]$\lambda$6363      &   1&  0       &     13.62     &     6.3\\
\ [N\,{\sc ii}]$\lambda$6548     &   1&  14.53   &     29.60     &     4.9\\
\ H$\alpha$\,$\lambda$6563       &   2&  13.60   &      --       &      --\\
\ [N\,{\sc ii}]6583     &   1&  14.53   &     29.60     &     4.9\\
\ [S\,{\sc ii}]$\lambda$6716     &   1&  10.36   &     23.34     &     3.2\\
\ [S\,{\sc ii}]$\lambda$6731     &   1&  10.36   &     23.34     &     3.6\\
\ [Ar\,{\sc iii}]$\lambda$7136   &   1&  27.63   &     40.74     &     6.7\\ 
\ [O\,{\sc ii}]$\lambda$7320     &   1&  13.62   &     35.12     &     6.8\\
\ [O\,{\sc ii}]$\lambda$7330     &   1&  13.62   &     35.12     &     6.8\\
 \hline
\end{tabular}
}
\end{center}
\footnotesize{\textbf{Notes}. The table shows the measured lines, the number of components used in the fitting, their ionization potentials (IP low) and the potential of the following ionic species (IP high), and the logarithm of the critical density.}\\
\label{tab:elenco_comp_a}
\end{table} 

\begin{figure}
 \centering
 \includegraphics[width=84 mm]{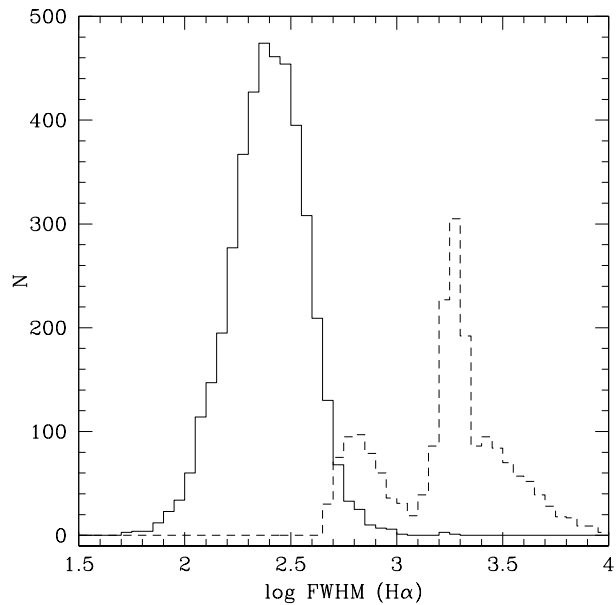}
 \caption{\small Distribution of $\rm FWHM (H\alpha)$ for the first component (solid line) and the second component (dashed line) of the emission-line.}
 \label{fig:FWHM_Ha_tot}
\end{figure}

\begin{figure}
 \centering
 \includegraphics[width=84 mm]{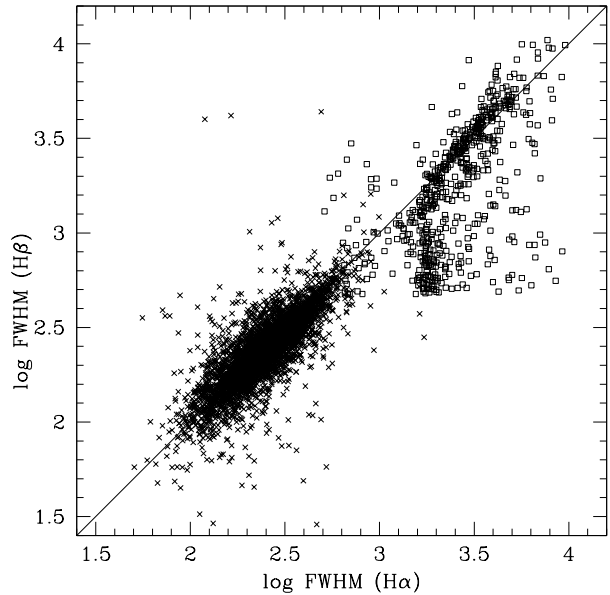}
 \caption{\small $\rm FWHM (H\beta)$ vs $\rm FWHM(H\alpha)$ for the 4254 measured spectra. Crosses are the first component, squares the second component.}
 \label{fig:FWHM}
\end{figure}

\begin{figure}
 \includegraphics[width=75 mm]{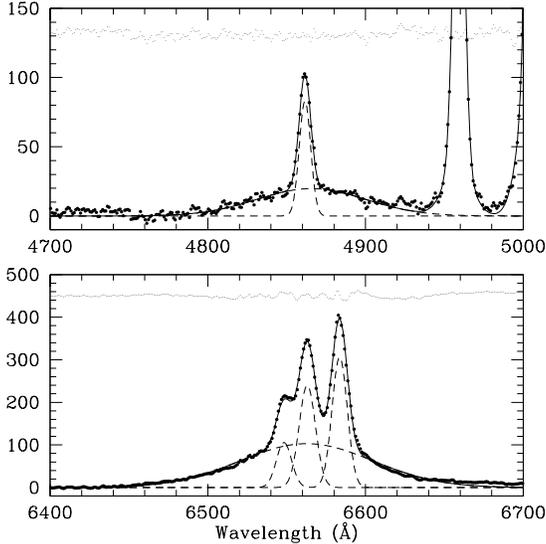}
\caption{\small Example of H$\beta$ (top) and H$\alpha$ (bottom) with a broad second component. The observed spectrum is in black dots, the components are in dashed line and the residuals in dotted line.}
\label{fig:hbetafit}
\end{figure}

\begin{figure}
 \centering
 \includegraphics[width=84 mm]{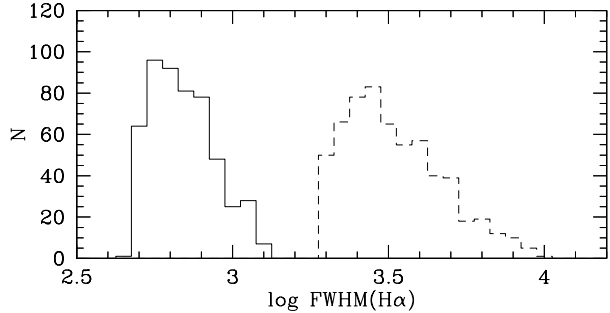}
 \caption{\small $\rm FWHM (H\alpha)$ distribution for the second component. The spectra showing H$\alpha$ second component with $\rm FWHM<1250 \,km s^{-1}$ are in solid line, while those having $\rm FWHM>2000 \,km s^{-1}$ are in dashed line.}
 \label{fig:FWHM_broad}
\end{figure}

The classification of Seyfert galaxies is based on the detection of the Balmer line broad components \citep{ost06}, but this criterion is subjective and it depends on the spectral quality. Additionally, visually classifying a large number of spectra is very difficult and extremely time-consuming.
In this work the criterion adopted for the spectral classification is based on the presence of a second component in the H$\alpha$ emission-line and on its FWHM. 
\citet{gel09} claimed that a good fitting of H$\alpha$ should be obtained using three components, a narrow, an intermediate and a broad one. However, it is in general difficult to find the corresponding components in H$\beta$, which is fundamental in our analysis. Therefore, in this work we decided to fit both H$\beta$ and H$\alpha$ with two components, one narrow and one broad.  
Out of the $4254$ measured spectra, $1938$ show a secondary H$\alpha$ component. In Fig.\,\ref{fig:FWHM_Ha_tot} we present the  $\rm FWHM (H{\alpha})$ distribution of both components after correction for instrumental width (R=1800). The $\rm FWHM (H{\alpha})$ distribution of the second component shows two peaks, one between $500$ and $1000$ km s$^{-1}$ and the other between $1500$ and $2200$ km s$^{-1}$, respectively. 
If the $\rm FWHM (H{\alpha})$ of the second component is lower than $2000$ km s$^{-1}$ the line is hidden in the H$\alpha$+[N\,{\sc ii}]$\lambda$6548,6584 feature \citep{gel09}. In this case, the SDSS spectral resolution is not sufficient to separate the true components. Therefore, since we could not be sure whether the second H$\alpha$ component is a real broad line or a mathematical result of the fit, we decided to exclude the broad components with $\rm{FWHM} < 2000$ km s$^{-1}$. 
In addition, we compared the widths of H$\alpha$ and H$\beta$ lines for all $4254$ measured spectra (Fig.\,\ref{fig:FWHM}). We noticed that the narrow components of both lines are well distributed around the 1:1 line, while the broad components are in good agreement only when the $\rm FWHM (H\alpha)>2000$ km s$^{-1}$.  
Below this value, the widths of the two emission-lines are totally different. In particular, the second component of H$\beta$ could not be a real broad line, or it could be poorly fitted, because too weak. 
When $\rm FWHM (H\alpha)<1250$ km s$^{-1}$ the second component of H$\beta$ disappears. 
Therefore, all the spectra with narrow $\rm H\alpha$ or with a second component narrower than $1250$ km s$^{-1}$ were classified as narrow-line spectra. The errors were taken into account, and the condition was fixed to $\rm FWHM (H\alpha)- 3\times \Delta FWHM (H\alpha)<1250 \,km s^{-1}$. The $\rm H\alpha$ flux was obtained as the sum of two fluxes when a second component was present. 
The spectra showing a second $\rm H\alpha$ component with $\rm FWHM (H\alpha)+ 3\times \Delta FWHM (H\alpha)>2000$ km s$^{-1}$ were classified as broad line spectra. In this second class we imposed an additional constraint: the reliability of the broad component was confirmed only when the height of its profile above the continuum at rest-frame 6531 and 6595 \AA\ was larger than $2\sigma_{\rm c}$, where $\sigma_{\rm c}$ is the root mean square of the continuum intensity. The factor $2$ was adopted based on visual tests on the detection of the broad component. Fig.\,\ref{fig:hbetafit} shows an example of broad-line spectra which satisfy the previously mentioned conditions. 
The distributions of the second component $\rm FWHM (H\alpha)$ of these groups are showed in Fig.\,\ref{fig:FWHM_broad}. 
Finally, $598$ objects were classified as broad emission-line galaxies and $2836$ as narrow emission-line galaxies. 
Then, we applied the VO diagnostic diagrams and the relations introduced by \citet{k06}. 
Only the objects belonging to the Seyfert region within $1\sigma$ error in the diagnostic ratios were considered. We obtained two final samples of $521$ Intermediate-type Seyfert (S-I) and $2153$ Seyfert 2 (S2) galaxies. 

\begin{figure}
 \centering
 \includegraphics[width=84 mm]{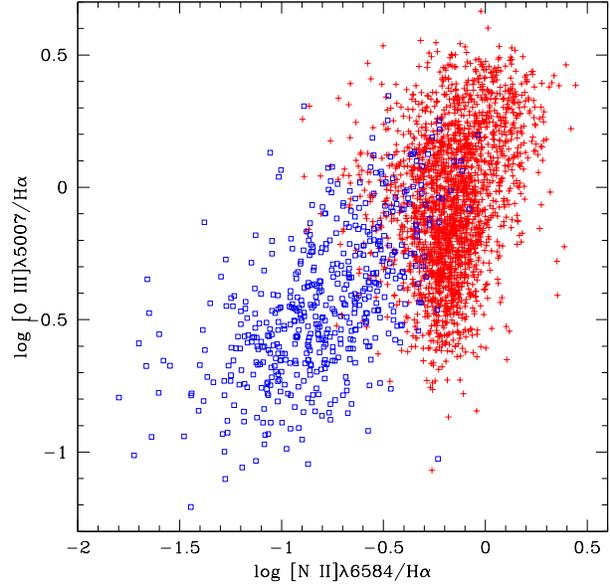}
 \caption{\small The logarithm of [N\,{\sc ii}]$\lambda$6584/H$\alpha$ ratio plotted against the logarithm of [O\,{\sc iii}]$\lambda$5007/H$\alpha$ ratio. S2 galaxies are indicated with red plus and S-I galaxies with blue squares.}
 \label{fig:O3HavsN2Ha}
\end{figure}

As a final check we plotted the $\log$([O\,{\sc iii}]$\lambda$5007/H$\alpha$) vs $\log$([N\,{\sc ii}]$\lambda$6584/H$\alpha$). \citet{gel09} found that this diagram is able to separate the S2 from the other Seyfert galaxies. They found that S2 galaxies have $\log$([O\,{\sc iii}]$\lambda$5007/H$\alpha)>0$ and $\log$([N\,{\sc ii}]$\lambda$6584/H$\alpha)>-0.5$, while our boundaries are $\log$([O\,{\sc iii}]$\lambda$5007/H$\alpha)>-0.6$ and $\log$([N\,{\sc ii}]$\lambda$6584/H$\alpha)>-0.3$ (Fig.\,\ref{fig:O3HavsN2Ha}). 
On the other hand, the correspondence is very good for the S-I galaxies. 
This discrepancy could be due to the fact that \citet{gel09} considered only spectra showing forbidden high-ionization line (FHIL), and so it is likely that [O\,{\sc iii}]$\lambda$5007/H$\alpha$ ratio is higher than the mean value we found in S2. 

\begin{figure}
 \centering
 \includegraphics[width=84 mm]{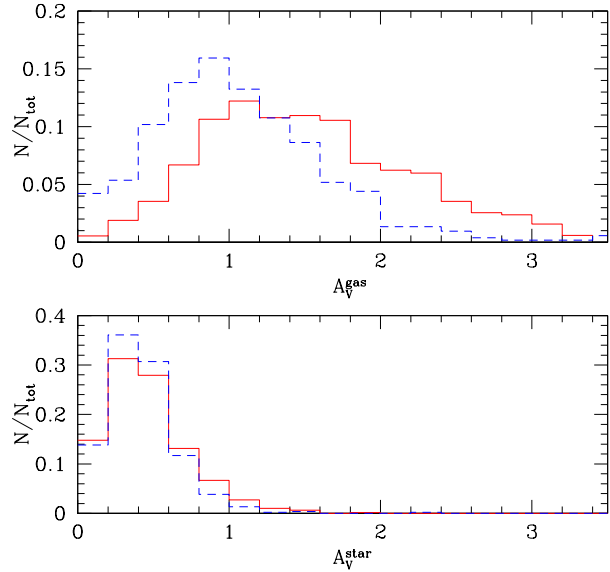}
 \caption{\small ${\rm A}(V)$ distribution for the samples: S2 (red solid line), and S-I (blue dashed line).}
 \label{fig:Av_cum}
\end{figure}

\begin{figure}
 \centering
 \includegraphics[width=84 mm]{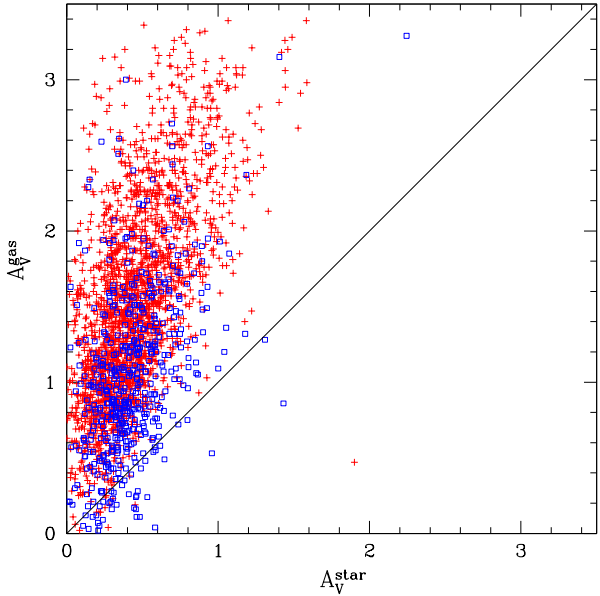}
 \caption{\small Stellar reddening (${\rm A}^{\rm star}_{V}$) versus gas reddening (${\rm A}^{\rm gas}_{V}$). Symbols are like in Fig.\,\ref{fig:O3HavsN2Ha}.}
 \label{fig:Av_Avstar}
\end{figure}

\section{Emission-line analysis}
In the following sections, we present the statistical analysis of the emission-line ratios. We looked for the correlations by means of the Spearman rank correlation coefficient (r$_s$) that is in general adopted when the distributions are a priori unknown. We used the two-tail test to verify the null hypothesis of each correlation, that is the probability that the correlation does not exist, and we assumed $\alpha=0.1$ per cent, as significance level.
We calculated the $t$ parameter of the Student distribution by following  \citep{2003psa..book.....W}, and we compared those values with tabulated critical t-values ($t_{crit}$) for a given number of degrees of freedom and a significance value. When $t>t_{crit}$, the correlation is significant at the chosen level and the null hypothesis can be rejected.

To compare samples, we used the Kolmogorov-Smirnov (K-S) test, to verify when two samples are drawn by the same population. When the observed parameter $\Delta$, that is the maximum difference between two distributions, is greater than the theoretical value D, the two samples are different with a significance level of 1 per cent.

\subsection{Internal reddening}\label{av}
The internal reddening correction was calculated using the parametrization introduced by CCM under the assumption H$\alpha$/H$\beta=3.1$ \citep{ost06}. 
Fig.\,\ref{fig:Av_cum} shows the distributions of the derived values of A$(V)$. The median values are $~1.4$ (90 per cent of the sample has A$(V)<2.5$) for the S2 and $~0.9$ (90 per cent of the sample has A$(V)<1.8$) for the S-I. Assuming a level of significance of 1 per cent, the Kolmogorov--Smirnov (KS) test gives 0.3, clearly different from the theoretical value of 0.08, indicating a statistical difference of A$(V)$ distribution between S2 and S-I. 
This result is expected in S-I and Seyfert 1 galaxies, because we are looking deeper inside the NLR, and the dust quantity is presumably lower because of the winds \citep{cks96} and the dust sublimation \citep{mul09}. 

The A$(V)$ values appear weakly correlated with stellar absorption (A$(V_{\star})$), evaluated by {\sc starlight} (Fig.\,\ref{fig:Av_Avstar}). Indeed the Spearman rank correlation coefficient is r$_s=0.55$, but the correlation is significant since $t=33.3 >> t_{crit}=3.3$ for N = 2557 targets (S2+S-I).

Our distribution is very similar to that obtained by \citet{gu06}, who measured the extinction of $65$ nearby Seyfert 2 nuclei from H$\gamma$/H$\beta$ ratio.
A$(V_{\star})$ values are distributed in a small interval and are very similar for both the samples. A large difference between the stellar and gas extinction was also found by \citet{cks94} and by \citet{cal97} in starburst galaxies. The same result was obtained by \citet{bennert06a} in a detailed analysis of the NLR in NGC 1386. This would suggest that most of the dust is mixed with the gas.

\subsection{Line intensity correlations}
\begin{figure*}
\centering
\includegraphics[width=8cm]{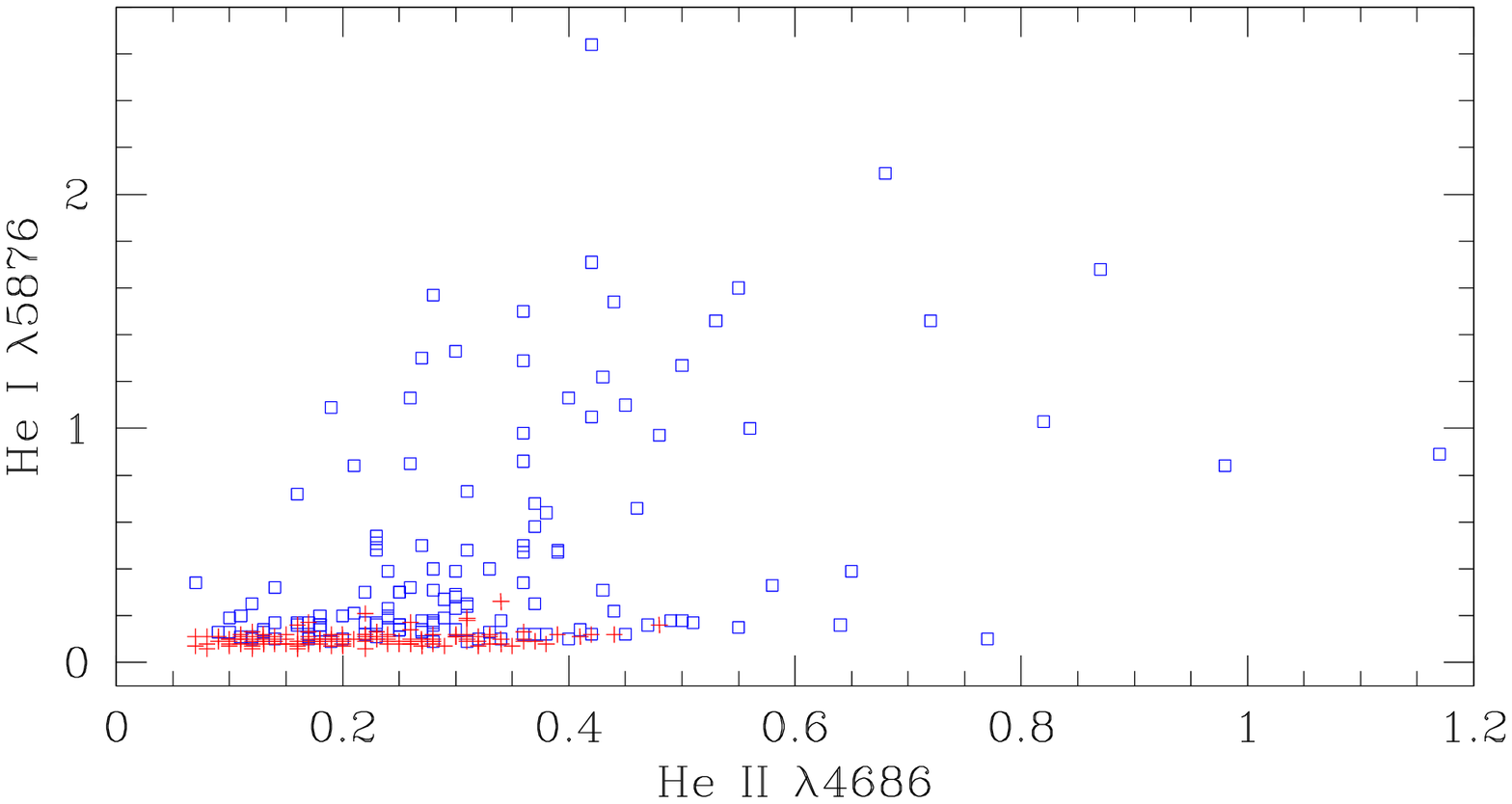}
\includegraphics[width=8cm]{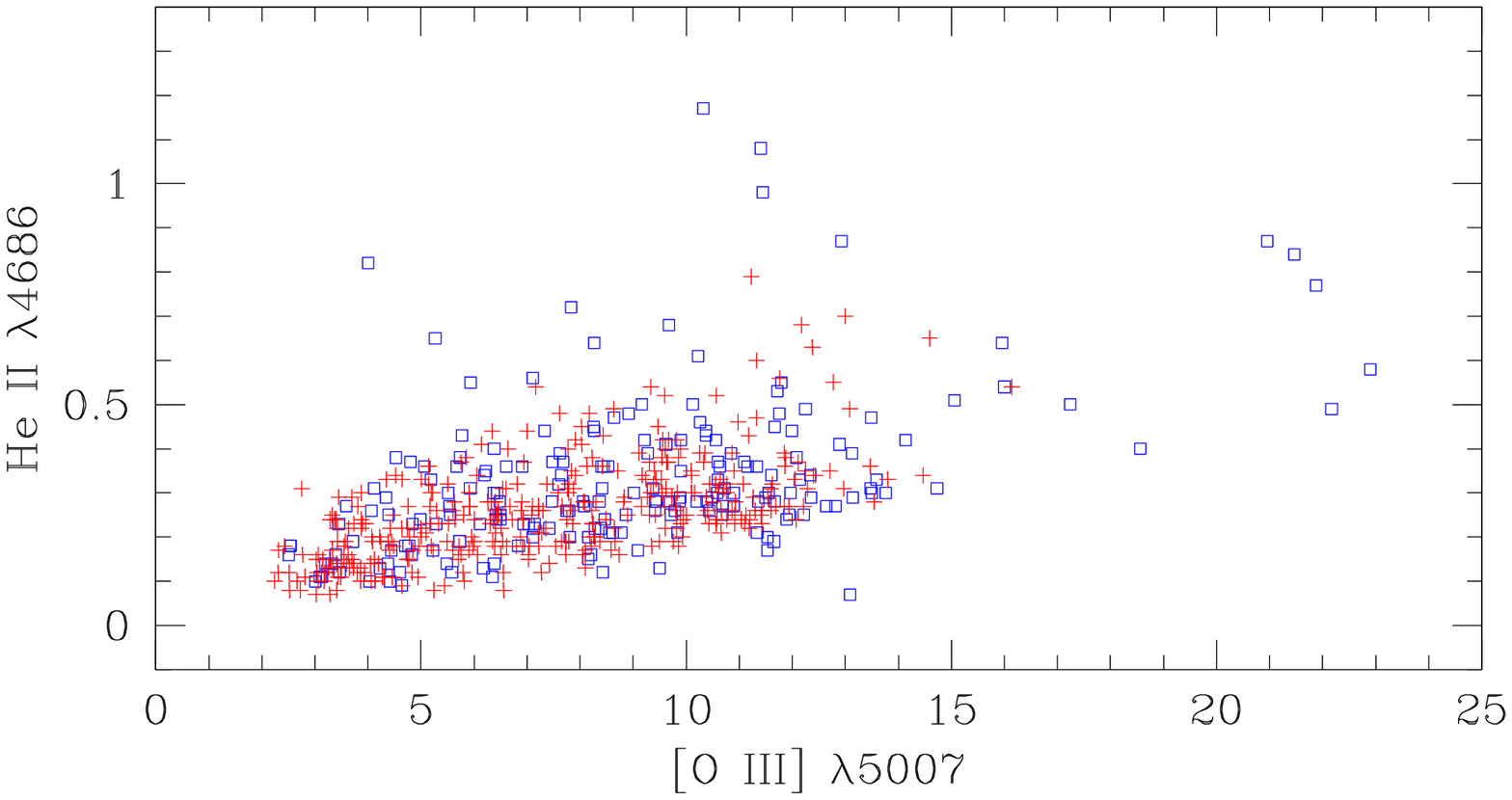}\\
\includegraphics[width=8cm]{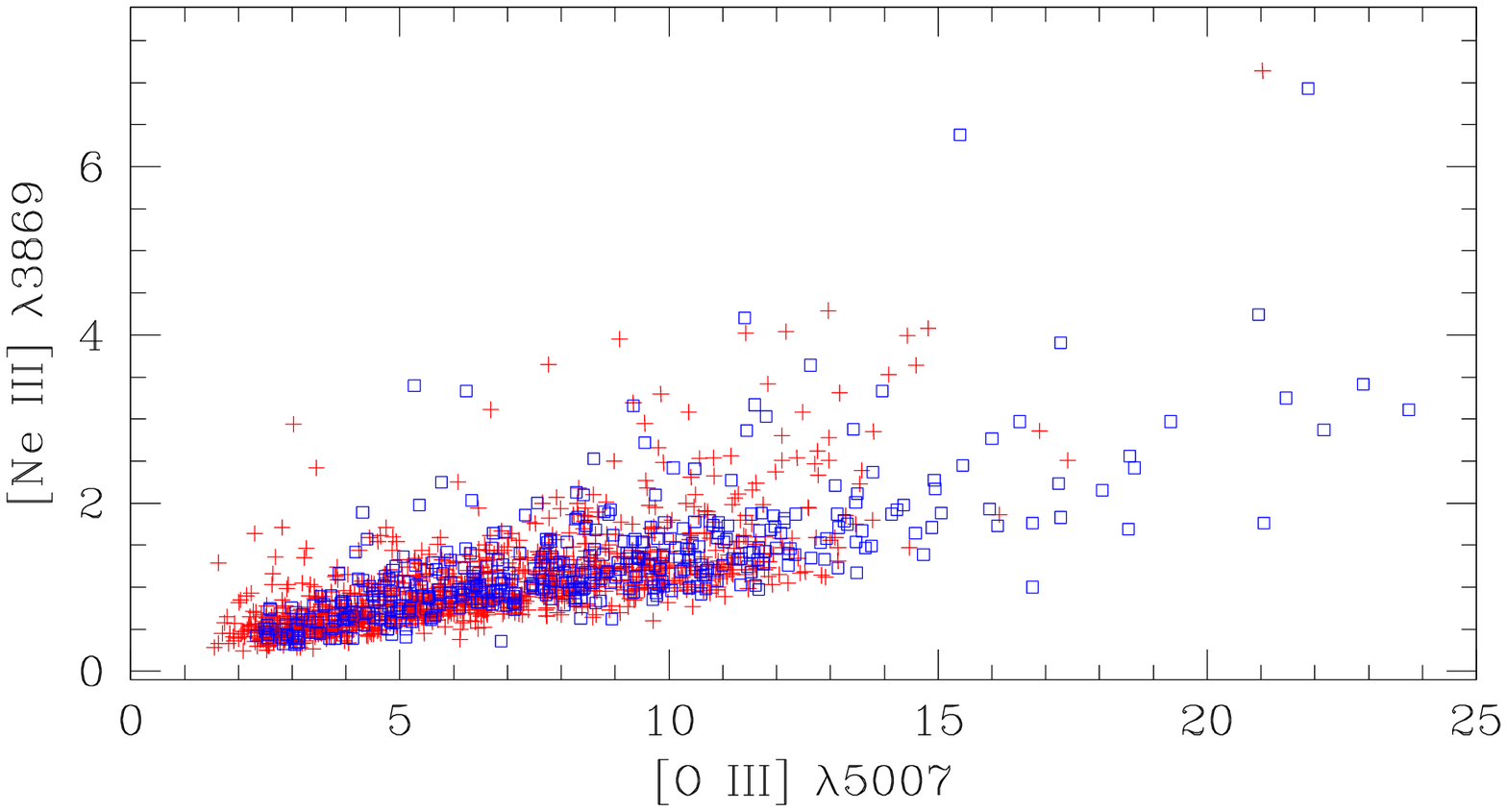}
\includegraphics[width=8cm]{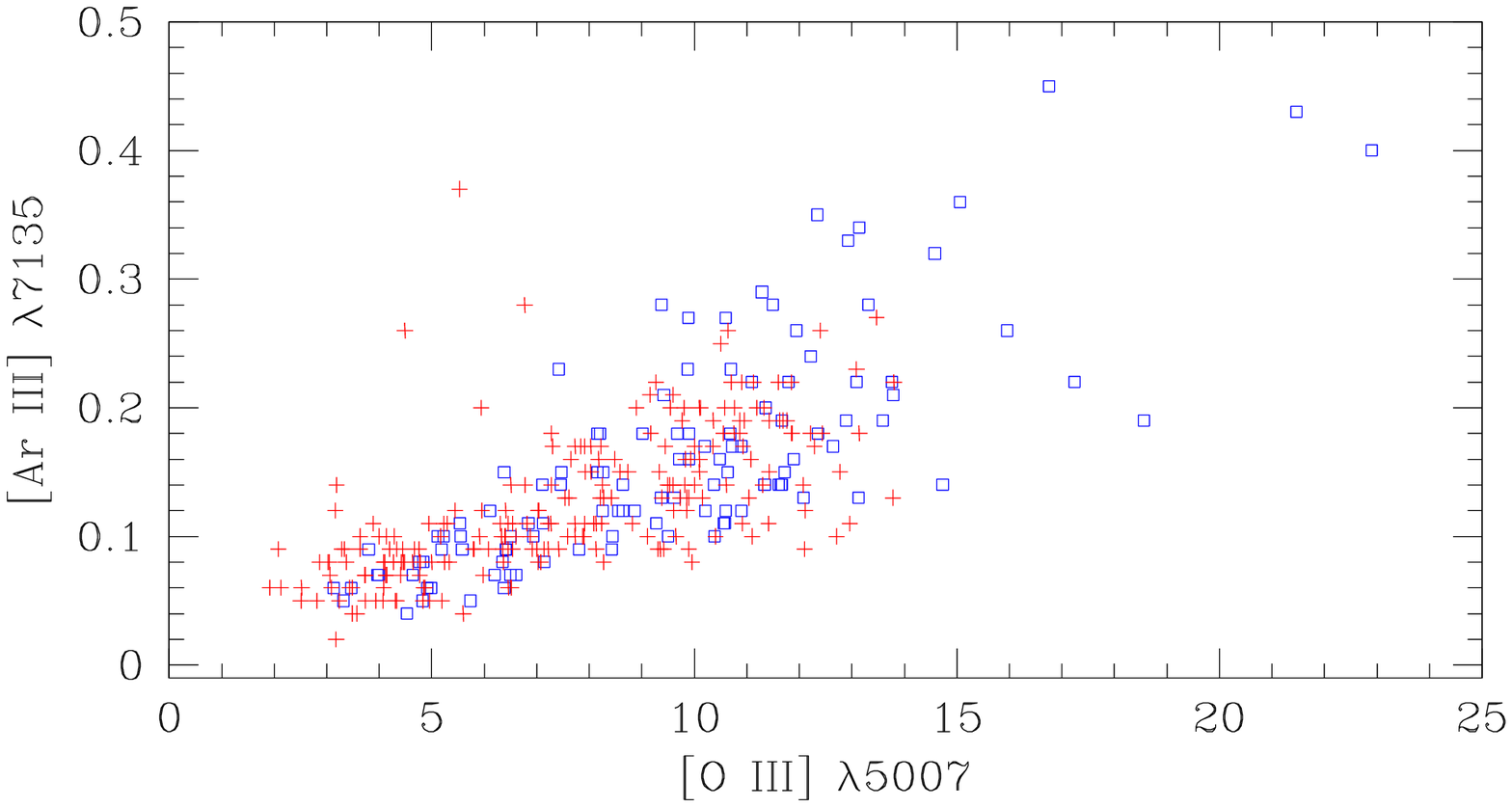}
\caption{\small Koski diagrams for high ionization lines (see text). The fluxes are relative to H$\beta$ and reddening corrected. Symbols are like in Fig.\,\ref{fig:O3HavsN2Ha}.}
\label{fig:k_5007}
\end{figure*}

\begin{figure*}
\centering
\includegraphics[width=8cm]{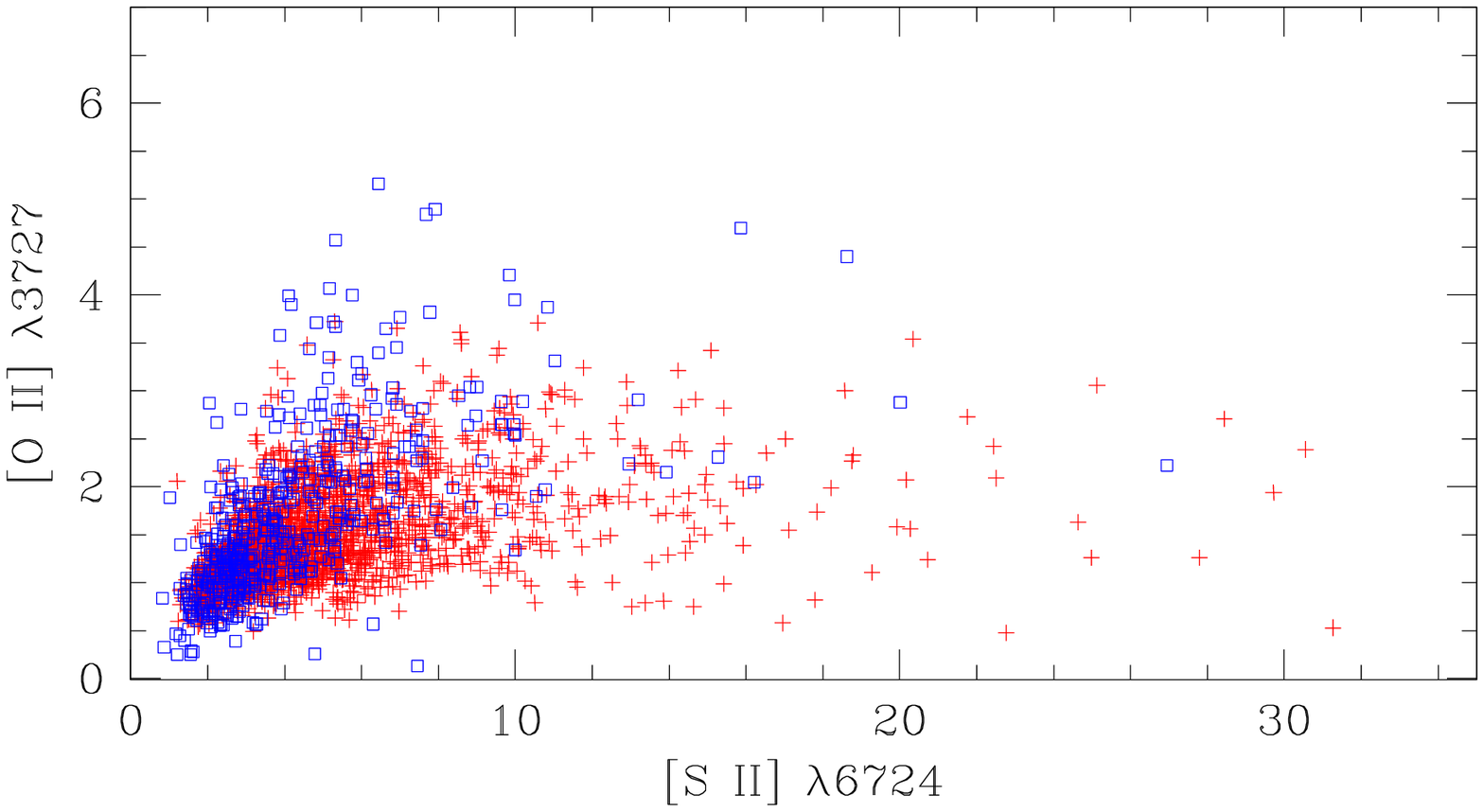}
\includegraphics[width=8cm]{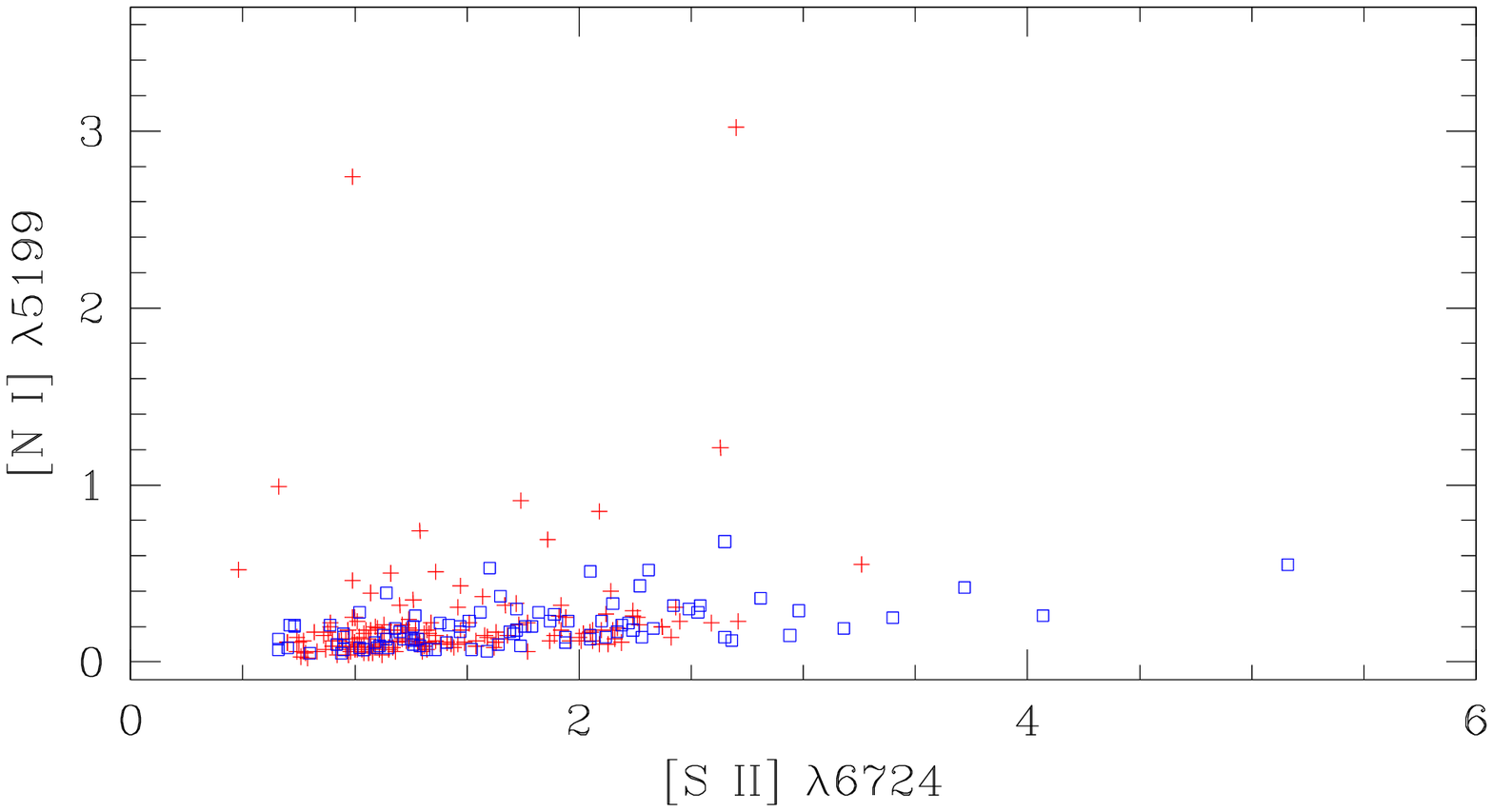}\\
\includegraphics[width=8cm]{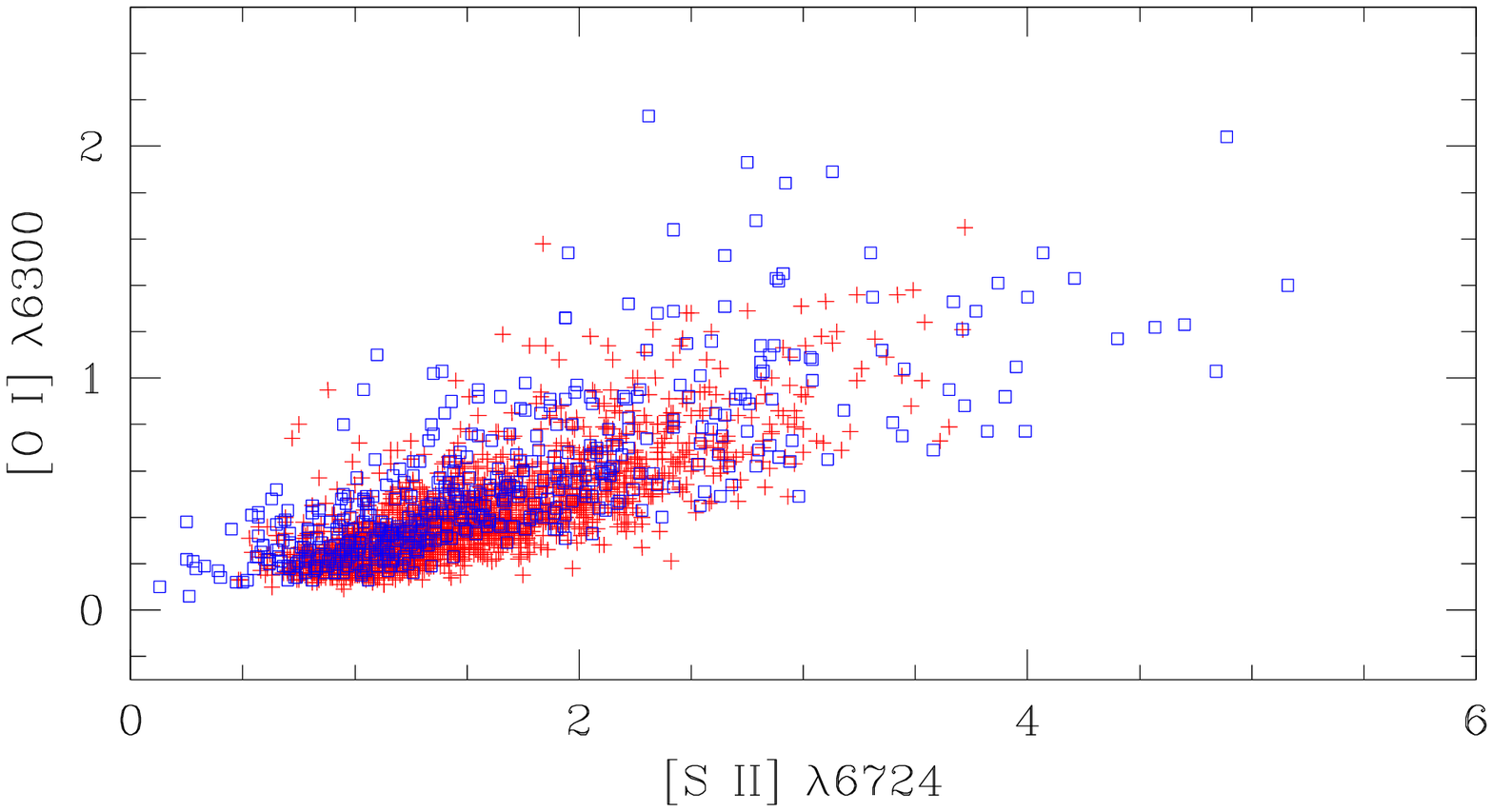}
\includegraphics[width=8cm]{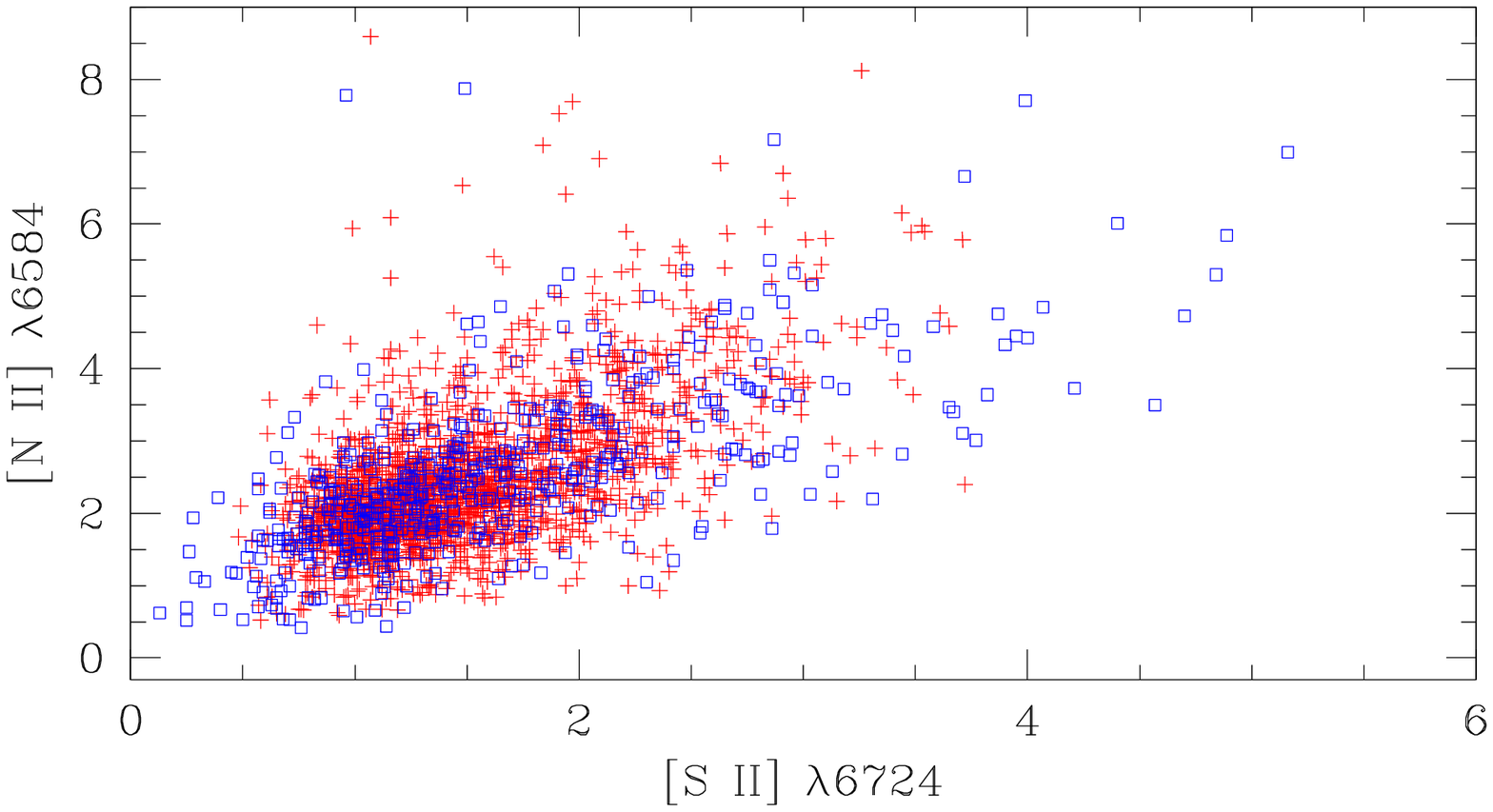}
\caption{\small Koski diagrams for low ionization lines (see text). The fluxes are relative to H$\beta$ and reddening corrected. Symbols are like in Fig.\,\ref{fig:O3HavsN2Ha}.}
\label{fig:k_6720}
\end{figure*}

The emission-line fluxes normalized to H$\beta$ and corrected for reddening were analysed following \citet{k78}. The emission-line ratios were grouped according to high and low ionization lines (Table \ref{tab:k_coeff}) and plotted versus the [O\,{\sc iii}]$\lambda$5007/H$\beta$ or [S\,{\sc ii}]$\lambda$6724/H$\beta$, respectively. As in \citet{k78} we added the He\,{\sc i} $\lambda$5876/H$\beta$ vs. He\,{\sc ii} $\lambda$4686/H$\beta$ diagram, in order to compare the helium recombination lines. These diagrams show the correlations between lines with similar ionization states (Figures \ref{fig:k_5007}, \ref{fig:k_6720}). 
Although the mean $\rm S/N$ of our spectra is lower than that of Koski, we found similar results by plotting a larger sample of objects.
\begin{table}
  \caption{Correlation coefficients between lines of similar ionization potential.}
\begin{center}
{\tiny
 \begin{tabular}{@{}|l|rcc|rcc|}
  \hline
  & & S2 & & & S-I &\\
& N & r$_s$ & $t$ & N & r$_s$ & $t$\\
\hline
 $\lambda$5876 vs. $\lambda$4686   & 154  & 0.17  & 2.1$^*$ & 133 & 0.38 & 4.7\\ 
 $\lambda$4686 vs. $\lambda$5007   & 354  & 0.60  & 14.1    & 199 & 0.49 & 7.9 \\
 $\lambda$3869 vs. $\lambda$5007   & 1126 & 0.77  & 40.5    & 431 & 0.77 & 25.0 \\
 $\lambda$7136 vs. $\lambda$5007   & 225  & 0.75  & 16.9    & 114 & 0.78 & 13.2 \\
 $\lambda$3727 vs. $\lambda$6724   & 2110 & 0.58  & 32.7    & 513 & 0.71 & 22.8 \\
 $\lambda$5199 vs. $\lambda$6724   & 165  & 0.38  & 5.2     & 88  & 0.50 & 5.4 \\
 $\lambda$6300 vs. $\lambda$6724   & 2105 & 0.76  & 53.6    & 508 & 0.80 & 30.0 \\
 $\lambda$6584 vs. $\lambda$6724   & 2100 & 0.51  & 27.2    & 509 & 0.69 & 21.5 \\
\hline
\end{tabular}
 }
 \end{center}
\footnotesize{\textbf{Note.} The asterisk indicates that the correlation is not significant.}
\label{tab:k_coeff}
\end{table}

Table \ref{tab:k_coeff} shows the number of targets (N), the Spearman rank correlation coefficient (r$_s$) and the parameter of the two-tail test ($t$). The critical value for the minimum number of targets, N = 88, is $t_{crit}=3.4$, therefore all correlations are significant at 0.1 per cent, but the first one for S2 galaxies, indicated by an asterisk.

[Ne\,{\sc iii}]$\lambda$3869 and [Ar\,{\sc iii}]$\lambda$7136 lines show a good correlation with the [O\,{\sc iii}]$\lambda$5007 line in both samples. He\,{\sc ii} $\lambda$4686 is well correlated only in S2 galaxies, while in S-I galaxies the correlation is significant, but weaker, likely because of the presence of the underlying broad component.    
Good or fairly good correlations exist between [O\,{\sc i}]$\lambda$6300 and [N\,{\sc ii}]$\lambda$6584 vs. [S\,{\sc ii}]$\lambda$6724, while [N\,{\sc i}]$\lambda$5199 vs. [S\,{\sc ii}]$\lambda$6724 is weak in both samples, probably because [N\,{\sc i}]$\lambda$5199 is very close to the Mg I $\lambda$5175 triplet and its measure can be strongly affected by the stellar continuum subtraction.
We found also good correlations between lines with similar critical density: [S\,{\sc ii}]$\lambda$6724 vs. [O\,{\sc ii}]$\lambda$3727, [Ar\,{\sc iii}]$\lambda$7136 vs. [O\,{\sc ii}]$\lambda$7325 (only for S2) and [Ne\,{\sc iii}]$\lambda$3869.

On the other hand, there is no correlation between the He\,{\sc i} and He\,{\sc ii} lines, and [Ar\,{\sc iii}]$\lambda$7136 vs. [O\,{\sc i}]$\lambda$6300 and [S\,{\sc ii}]$\lambda$4074. These last lines have similar critical densities, but very different ionization potentials, supporting the hypothesis of two independent ionization zones. 
We do not find any correlation between the [O\,{\sc iii}]$\lambda$5007 line and the high ionization lines [Fe\,{\sc vii}]$\lambda$5721,6087 and [Ar\,{\sc iv}]$\lambda$4711,4740. 
The [Fe\,{\sc vii}] lines have an higher critical density compared to the [O\,{\sc iii}] lines, while the [Ar\,{\sc iv}] lines have higher ionization potential and lower critical density. Therefore, these lines could be formed in different regions compared with the [O\,{\sc iii}]$\lambda$5007 line. 
[Fe\,{\sc vii}]$\lambda$5721,6087 were detected in few spectra, but the percentage increases from S2 to S-I (see Table \ref{tab:fe7_O3_a}). These lines are formed in the coronal-line region (CLR) and are very important in the analysis of the NLR structure. \citet{mt98} claimed that in the context of the Unified Model three kinds of CLR exist: those associated with the torus, with the NLR and with the ENLR. The condition to form the coronal lines is a dust free medium with densities of 10$^2$--10$^{8.3}$ cm$^{-3}$ \citep{ferguson97}. If the CLR is far away from the source, it is likely to have a very low density, $\sim 1$ cm$^ {-3}$ \citep{kor89}. 

\begin{table}
\caption{Number of spectra with [Fe\,{\sc vii}] and [O\,{\sc iii}]$\lambda$4363 lines and their percentage with respect to the samples.}
\begin{center}
{\tiny
\begin{tabular}{|l|l|l|l|}
\hline
samples & [Fe\,{\sc vii}]6086 & [O\,{\sc iii}]$\lambda$4363 & [O\,{\sc iii}]$\lambda$4363+[Fe\,{\sc vii}]$\lambda$6087 \\ 
\hline
S2 &  96        (4 per cent)   & 86        (4 per cent) & 38           (2 per cent) \\ 
S-I &  97        (19 per cent) & 186     (36 per cent) & 73          (14 per cent)\\
\hline
\end{tabular}
}
\end{center}
\label{tab:fe7_O3_a}
\end{table}

\begin{figure}
 \centering
 \includegraphics[width=7cm]{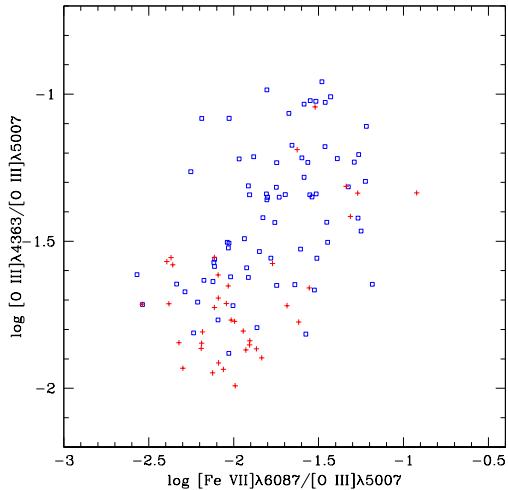} 
 \caption{\small The logarithm of the [Fe\,{\sc vii}]$\lambda$6087/[O\,{\sc iii}]$\lambda$5007 ratio is plotted against the logarithm of the [O\,{\sc iii}]$\lambda$4363/[O\,{\sc iii}]$\lambda$5007 ratio for the two samples. Symbols are like in Fig.\,\ref{fig:O3HavsN2Ha}.}
\label{fig:fe7_4363}
 \end{figure}

Since the critical density of the [Fe\,{\sc vii}]$\lambda$6087 line is very similar to that of the [O\,{\sc iii}]$\lambda$4363 line, we followed \citet{nag01} and analyzed a possible correlation between these two lines. About 50 per cent of the spectra where the [O\,{\sc iii}]$\lambda$4363 line is detected shows the [Fe\,{\sc vii}]$\lambda$6087 line.
In particular, we calculated [Fe\,{\sc vii}]$\lambda$6087/[O\,{\sc iii}]$\lambda$5007 and [O\,{\sc iii}]$\lambda$4363/[O\,{\sc iii}]$\lambda$5007 ratios. We did not find correlation for S2, r$_s$=0.28  ($t=1.8 < t_{crit}=3.6$, N=38) and a weak but significant correlation for S-I, r$_s$=0.47 ($t=4.5 > t_{crit}=3.4$, N=73) (Fig.\,\ref{fig:fe7_4363}). 
The low value of the correlation index in the S-I sample could be due to larger errors in the measurements of the flux of [O\,{\sc iii}]$\lambda$4363, because of the presence of a non-negligible broad H$\gamma$ component. 
It is interesting to note that the distribution extends to higher values in the S-I case. Indeed, if the temperature increases and the density decreases, then the [O\,{\sc iii}]$\lambda$4363/[O\,{\sc iii}]$\lambda$5007 ratio increases \citep{ost06}. This is consistent with the assumption that the coronal lines are forming in a medium with low or intermediate density and high temperature.
The same critical density is not a sufficient condition to support the idea that these lines are emitted in the same region. Their ionization potentials are very different, therefore it is likely that these lines are in general emitted in different regions and in some cases in partially overlapped regions.

\section{Physical characteristics}\label{physic}

\subsection{Densities and Temperatures}
\begin{figure}
 \centering
 \includegraphics[width=7cm]{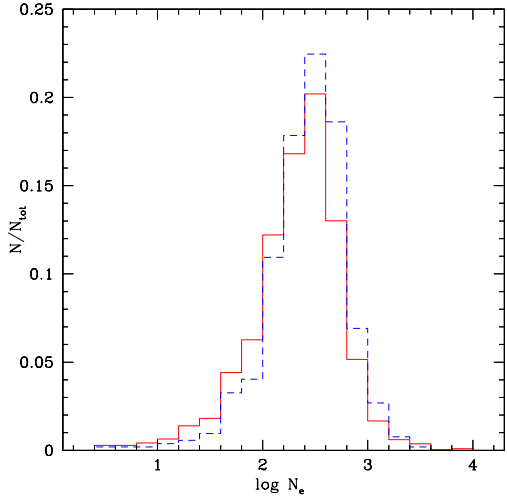}
 \caption{\small Density distributions calculated from RS2 (see text). 
Colours and lines are like in Fig.\,\ref{fig:Av_cum}.}
 \label{fig:S2_den}
\end{figure}

\begin{figure}
 \centering
 \includegraphics[width=7cm]{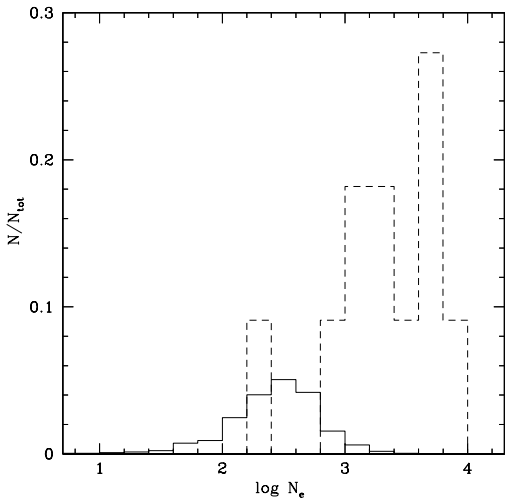}
 \caption{\small Density distributions calculated from RS2 (thick line) and RAr4 (dashed line).}
 \label{fig:S2Ar4_den}
\end{figure}

The values of density and temperature were calculated by means of the IRAF task {\sc temden}. 
The line ratios used to determine the densities were [S\,{\sc ii}]$\lambda$6716/6731 (hereafter RS2) and [Ar\,{\sc iv}]$\lambda$4711/4740 (hereafter RAr4), while the ratios used for the temperature were [O\,{\sc iii}](4959+5007)/4363 (hereafter RO3), [S\,{\sc ii}](6716+6731)/(4068+4076) (hereafter RS2t), [O\,{\sc ii}]$\lambda$3727/$\lambda$7325 (hereafter RO2). A temperature of 10000 K was assumed for the density determination. 
Fig.\,\ref{fig:S2_den} shows the distribution of the densities calculated by means of RS2. This ratio was measurable in all spectra, but reasonable values ($\rm RS2>0.29$, theoretical lower limit) and so reliable estimates of the density were possible in 2316 out of the 2674 objects, 1846 S2 and 470 S-I (87 per cent of the total sample). 
The median value is $N_{\rm e} \sim 250$ cm$^{-3}$, most of objects ($\sim2300$ galaxies) show values lower than $500$ cm$^{-3}$, while only in 97 cases we observe densities higher than 1000 cm$^{-3}$.  Only in $11$ objects it was possible to estimate the density with argon lines, $7$ S2 and $4$ S-I. The obtained values are of the order of $10^{3}$ cm$^{-3}$ (Fig.\,\ref{fig:S2Ar4_den}). This is not unexpected, since the [Ar\,{\sc iv}] transitions are characterized by higher critical density values than those of the [S\,{\sc ii}] lines, and therefore they are probably emitted by gas in different physical conditions. Interestingly, the [O\,{\sc iii}]$\lambda$4363 transition, which shares a similar ionization potential, has an even higher critical density, and this means that the gas emitting this line is also characterized by a larger electron density, likely around $10^{4}$ cm$^{-3}$ or even more.
Moreover, both lines [Ar\,{\sc iv}]$\lambda$4711 and [Ar\,{\sc iv}]$\lambda$4740 are generally weak, also after reddening correction, showing median intensities normalized to H$\beta$ of $0.07$ and $0.09$, respectively. Since most of the spectra, where they could be detected and measured, have S/N ratio between $20$ and $40$, we can in principle conclude that these lines are visible only in few galaxies, because spectra with high S/N ratio in the continuum are mandatory. Anyway, this could be only a necessary but not sufficient condition. Indeed, 66 per cent of our sample has $\rm S/N >20$. Moreover, one can speculate for example that  [O\,{\sc iii}]$\lambda$4363 has a median intensity only twice higher than [Ar\,{\sc iv}]$\lambda$4711 and [Ar\,{\sc iv}]$\lambda$4740, and it was detected in $272$ spectra having S/N ratio between $15$ and $40$, notwithstanding the fact that oxygen is about $200$ times more abundant than argon, assuming solar abundances. Therefore, it is also possible that the gas of the NLR is made essentially by a low and a high density medium, and that the second one has the necessary electron density to collisionally populate the oxygen auroral line and simultaneously suppress the [Ar\,{\sc iv}] lines.
\begin {table}
\begin{center}
\caption{Median values of electron temperature for the samples. The meaning of $D$ and $\Delta$ is explained at the beginning of Section 4.}
\begin{tabular}{|l|l|l|l|l|l|l|}
\hline
  \multicolumn{1}{|c|}{temperature} &
  \multicolumn{1}{|c|}{S2}&
  \multicolumn{1}{|c|}{S-I}&
  \multicolumn{1}{c|}{N S2} &
  \multicolumn{1}{c|}{N S-I} &
  \multicolumn{1}{c|}{D} &
  \multicolumn{1}{c|}{$\Delta$} \\
\hline
  TO2 & 12000 & 10000 & 68 & 13 & 0.49 & 0.14\\
  TS2 & 11000 & 15000 & 63 & 53 & 0.30 & 0.30\\
  TO3 & 17500 & 25000 & 84 & 170& 0.21 & 0.42\\
\hline
\end{tabular}
\label{tab:Te_median}
\end{center}
\end{table}

\begin{figure}
 \centering
 \includegraphics[width=7cm]{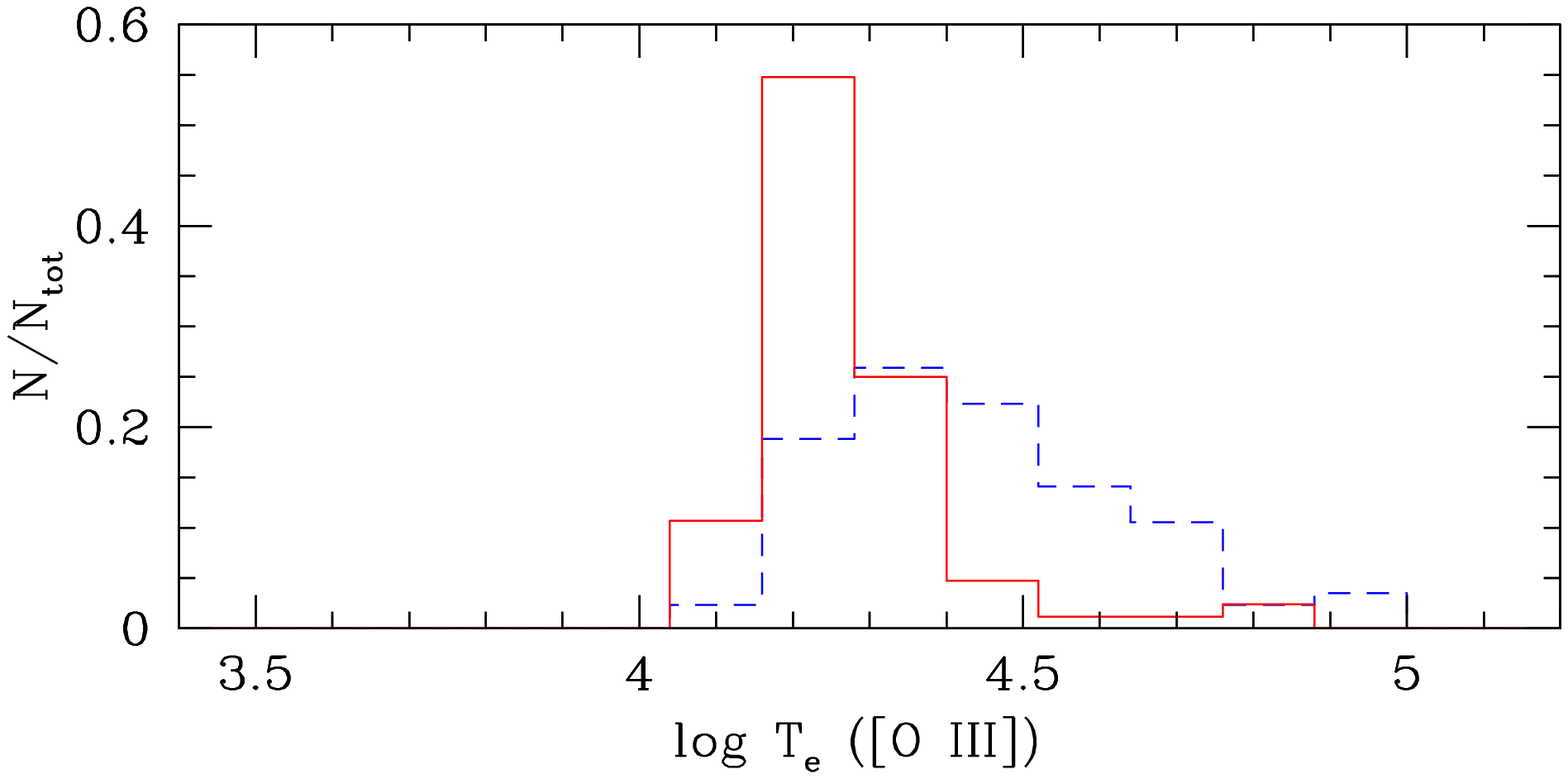}
 \includegraphics[width=7cm]{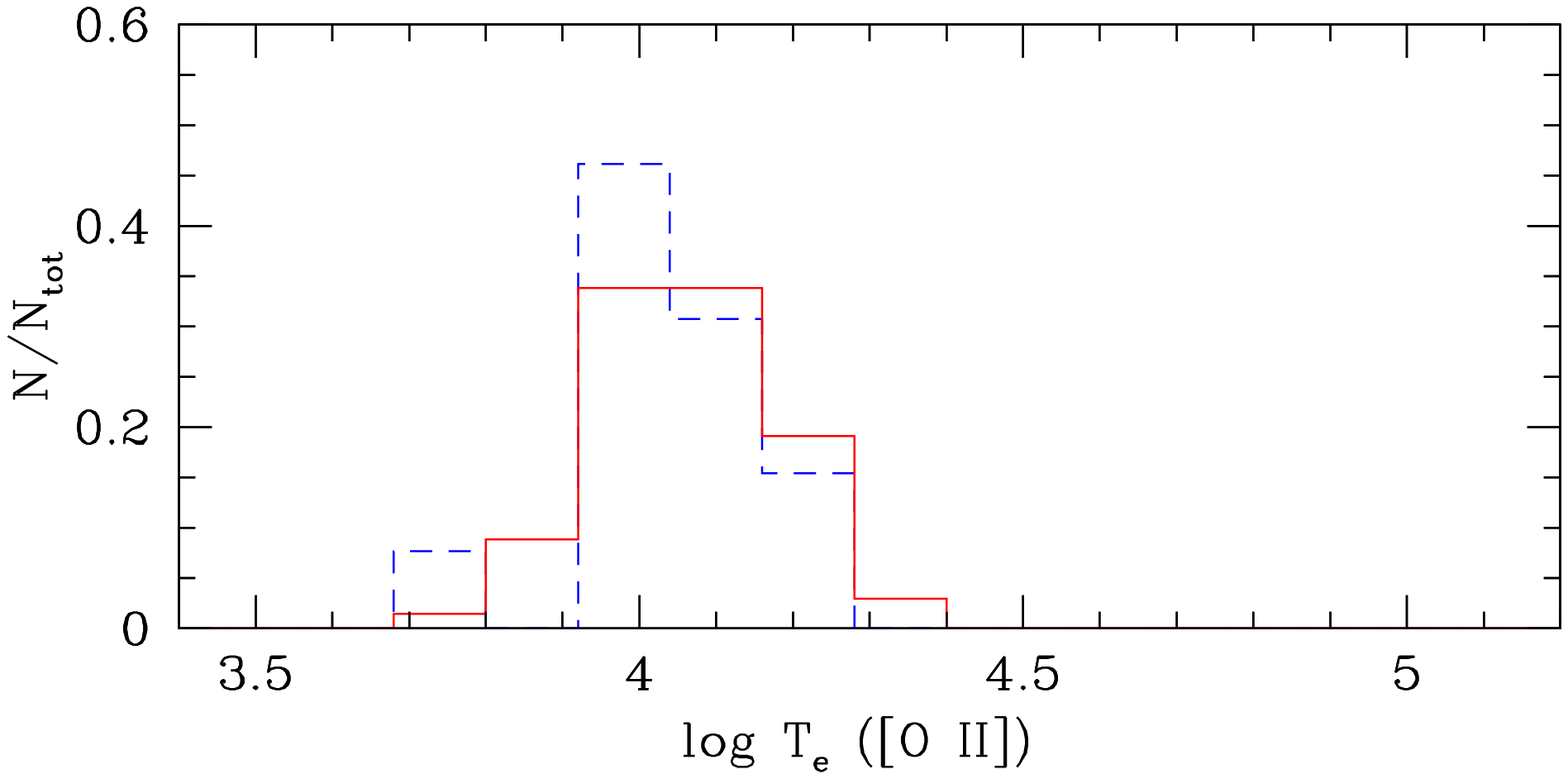}
 \includegraphics[width=7cm]{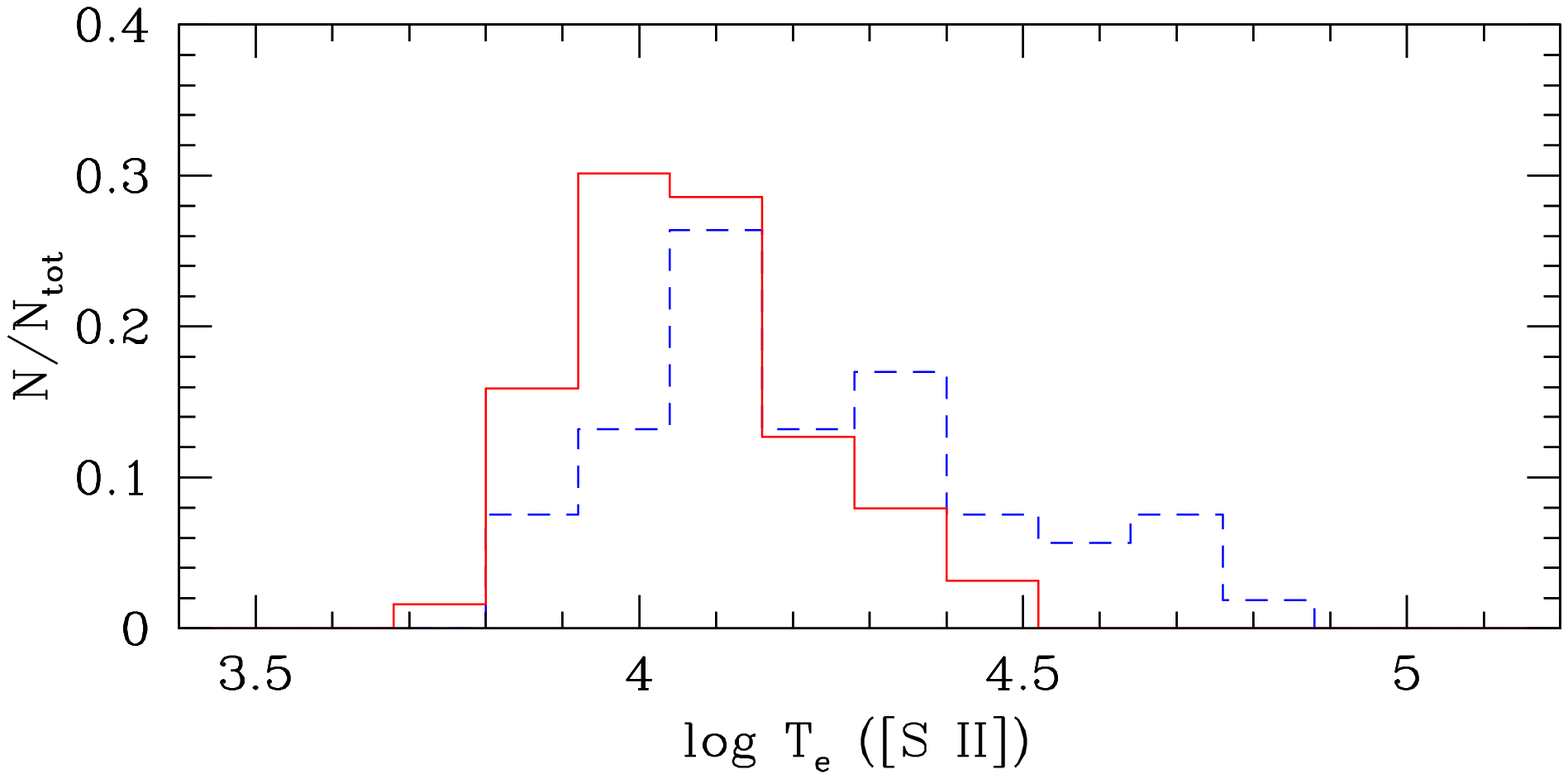}
 \caption{\small Distribution of temperature in logarithmic scale, measured by means of different line ratios: $T_{\rm e}$([O\,{\sc iii}]) (top), $T_{\rm e}$([O\,{\sc ii}]) (middle),  $T_{\rm e}$([S\,{\sc ii}]) (bottom). Colours and lines are like in Fig.\,\ref{fig:Av_cum}.}
 \label{fig:Te}
\end{figure}

The estimate of temperature can be difficult because sometimes the lines involved are weak or even very weak. Anyway, reliable estimates were obtained with RO3 in 254 objects, 84 S2 and 170 S-I, with RO2 in 81 objects, 68 S2, 13 S-I, and finally with RS2t in 116 reliable measures, 63 S2 and 53 S-I. 
Fig.\,\ref{fig:Te} shows the distributions of the so determined temperatures for both samples while the median values are reported in Table \ref{tab:Te_median}.
We used the electron density obtained with RS2 as input to {\sc temden}. Through a two-sample Kolmogorov--Smirnov test (see Table \ref{tab:Te_median}) it appears that the distributions of the temperatures derived from [O\,{\sc ii}] and [S\,{\sc ii}] are very similar for S2. 
On the other hand, the temperatures obtained from [O\,{\sc iii}] have a completely different distribution. In principle, this result could be indicative of the presence of ion stratification according to the ionization potentials. In contrast, it could be simply the effect of the use of $N_{\rm e}$([S\,{\sc ii}]), which indicate too low electron density values to collisionally pump the [O\,{\sc iii}]$\lambda$4363. The use of $N_{\rm e}$([Ar\,{\sc iv}]) should be more correct, nevertheless the critical densities of [Ar\,{\sc iv}] lines are lower than that of [O\,{\sc iii}] $\lambda$4363, and this suggests that $N_{\rm e}$([Ar\,{\sc iv}])=10$^3$--10$^4$ cm$^{-3}$ could be a lower limit for the density of the gas emitting the [O\,{\sc iii}] auroral line. 

By assuming that photoionization from the active nucleus is the main physical mechanism taking place in the NLR, we can conclude that [O\,{\sc iii}]$\lambda$4959,5007 are likely emitted both by a low density medium at higher temperature reflecting ion stratification, and by high density clouds with high ionization, where the physical conditions are such as to allow the [O\,{\sc iii}]$\lambda$4363 to form and become detectable and measurable.
The high values of temperatures measured with [O\,{\sc iii}] lines in the S-I sample, seems to indicate that we are observing the part of the NLR closer to the AGN. However, we stress that this result could be an effect of non perfect subtraction of H$\gamma$ broad component.

\subsection{Ionization parameter}
\begin{figure}
 \centering
 \includegraphics[width=84 mm]{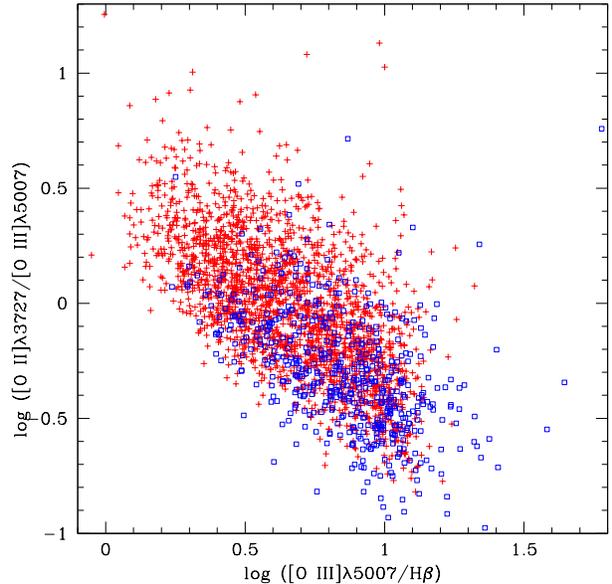}
 \caption{\small The logarithm of the [O\,{\sc iii}]$\lambda$5007/H$\beta$ ratio is plotted against the logarithm of the [O\,{\sc ii}]$\lambda$3727/[O\,{\sc iii}]$\lambda$5007 ratio. Symbols are like in Fig.\,\ref{fig:O3HavsN2Ha}.}
 \label{fig:U_ratios}
\end{figure}

\begin{figure}
 \centering
\includegraphics[width=8cm]{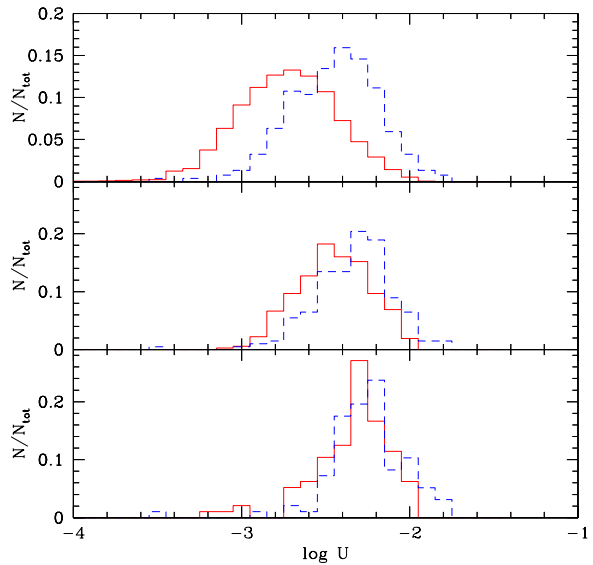}
 \caption{\small Ionization parameter distributions. Top: S2 sample is indicated in red solid line and the S-I sample in blue dashed line. Middle: the distributions taking into account only the spectra with the He\,{\sc ii} $\lambda$4686 measured, bottom: the distributions taking into account only the spectra with the [Fe\,{\sc vii}]$\lambda$6087 measured.}
 \label{fig:U_dist}
\end{figure}

\begin{figure}
 \centering
\includegraphics[width=84 mm]{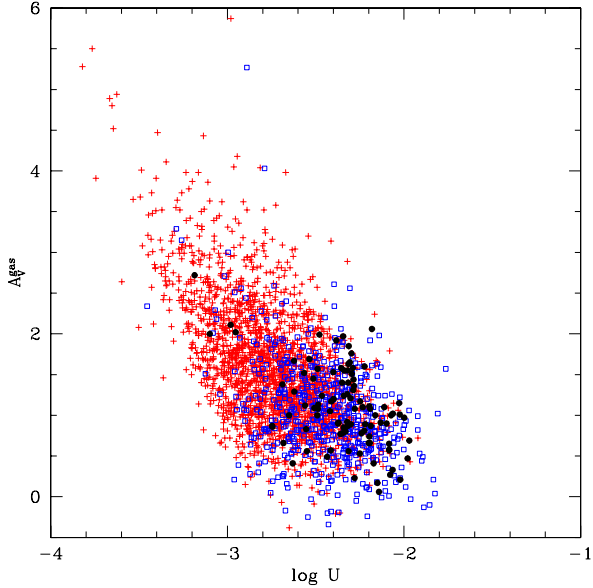}
 \caption{\small The logarithm of the ionization parameter is plotted against the reddening derived from emission-lines. Symbols are like in Fig.\,\ref{fig:Av_cum}. Black filled circles are Seyfert 2 showing [Fe\,{\sc vii}].}
 \label{fig:U_Av}
\end{figure}

The ionization level is defined by the ionization parameter $U$, which is the ratio between the flux of ionizing photons and the hydrogen density,
\begin{equation}
U= \frac{Q_{\rm ion}}{{\rm c}\,r^{2}\,N_{\rm H}}
\end{equation}
where $Q_{\rm ion}$ is the number of ionizing photons per second emitted by the source.
The ionization parameter could be evaluated using two ratios, [O\,{\sc iii}]$\lambda$5007/H$\beta$, as suggested by \citet{cohen83}, and the most recent ratio [O\,{\sc ii}]$\lambda$3727/[O\,{\sc iii}]$\lambda$5007 \citep{ks97}. We plotted the [O\,{\sc ii}]$\lambda$3727/[O\,{\sc iii}]$\lambda$5007 vs [O\,{\sc iii}]$\lambda$5007/H$\beta$ diagram (Fig.\,\ref{fig:U_ratios}).
The trend is clear in both the samples: r$_s=-0.65$ ($t=39.7$, N=2153) for S2, $\rm r_s=-0.56$ ($t=15.4$, N=521) for S-I. 
Cohen \& Osterbrock (1981) defined high ionization when [O\,{\sc iii}]$\lambda$5007/H$\beta>10$. They analyzed different line ratios in order to define a non-arbitrary way to classify high and low ionization level. In their fig. 2, [Fe\,{\sc vii}]$\lambda$6087/H$\beta$ vs. [O\,{\sc iii}]$\lambda$5007/H$\beta$ diagram, they found that [Fe\,{\sc vii}]$\lambda$6087/H$\beta$ shows a cutoff: [Fe\,{\sc vii}]$\lambda$6087 is not observed, when [O\,{\sc iii}]$\lambda$5007/H$\beta<10$. 
From Fig.\,\ref{fig:U_ratios} we can estimate that this limit corresponds to [O\,{\sc ii}]$\lambda$3727/[O\,{\sc iii}]$\lambda$5007 lower than 0.5 and 0.4 respectively for S2 and S-I. 
It corresponds to a log(U) greater than -2.5 and -2.4 respectively for S2 and S-I, by assuming the relation between the [O\,{\sc ii}]$\lambda$3727/[O\,{\sc iii}]$\lambda$5007 ratio and the ionization parameter U introduced by \citet{penston90}:
\begin{equation}
\log U=-2.74-\log(I_{3727}/I_{5007})
\label{eq:eq4}.
\end{equation}
Using Eq.\,\ref{eq:eq4} we estimated the ionization parameter for both the samples, and we applied the KS test subdividing each sample in two sub-samples of spectra (Table \ref{tab:U_KS}), the first one showing the He\,{\sc ii} $\lambda$4686 emission-line and the second one showing the [Fe\,{\sc vii}]$\lambda$6087 emission line. There is a significant difference between S2 and S-I galaxies when all the spectra are taken into account. It is interesting to point out that this difference becomes less significant when we compare the objects emitting He\,{\sc ii} $\lambda$4686 and disappears when the [Fe\,{\sc vii}]$\lambda$6087 is taken into account (Fig.\,\ref{fig:U_dist}). 
The ionization parameter depends on the luminosity and is inversely proportional to density and distance from the source. The different distributions suggest that one or more of these parameters change. If the luminosity changes, then S-I and S2 are intrinsically different objects and the Unified Model fails. On the other hand, since the average electron density seems to be the same in the two samples (see Section \ref{physic}), changes in the ionization parameter could be a distance effect. The ionization is highly stratified in the NLR and in S-I galaxies we are able to observe closer to the active nucleus, therefore we can speculate that S2 galaxies showing He\,{\sc ii} and [Fe\,{\sc vii}] have a wider aperture angle of the torus, which is thinner in the direction orthogonal to the line of sight. 
Indeed, by plotting $\log U$ vs. A$(V)$ we can observe that S2 galaxies showing [Fe\,{\sc vii}] are characterized by a relatively low extinction, with values similar to those observerd in S-I galaxies (Fig.\,\ref{fig:U_Av}). 
\begin{table}
\caption{Two-sample Kolmogorov--Smirnov test concerning the ionization potential distributions with 1 per cent of significance.}
\begin{center}
{\small
\begin{tabular}{|r|r|r|}
\hline
  \multicolumn{1}{|c|}{samples} &
  \multicolumn{1}{c|}{D} &
  \multicolumn{1}{c|}{${\Delta}$} \\
\hline
  S2 $\leftrightarrow$ S-I      & 0.08  & 0.38\\
  (S2 $\leftrightarrow$ S-I)$_{\lambda4686}$ & 0.14  & 0.25\\
  (S2 $\leftrightarrow$ S-I)$_{\lambda6087}$ & 0.23  & 0.21\\
\hline
\end{tabular}
}
\end{center}
\footnotesize{\textbf{Notes.} (S2 $\leftrightarrow$ S-I)$_{\lambda4686}$ and (S2 $\leftrightarrow$ S-I)$_{\lambda6087}$ indicates the comparison between the samples when the spectra show respectively the He\,{\sc ii} $\lambda$4686 and [Fe\,{\sc vii}]{$\lambda6087$} emission-lines.}\\
\label{tab:U_KS}
\end{table}

\subsection{Ionized gas mass and radius}
\citet{ho09} suggests that the mass budget of the NLR can be accounted by mass loss from evolved stars. From  H$\alpha$ (or H$\beta$) luminosity it is possible to determine the ionized gas mass \citep{ost06}. The author found $M_{\rm NLR}\sim3\times10^{4}\,{\rm M}_{\odot}$ in a region of $200\times400$ pc. This value comes from the population statistics of the Palomar survey assembled in \citet{ho03} and it probably underestimates the mass by no more than a factor of 2. 
Comparing this mass with the amount of material shed by evolved stars, determined by the models of \citet{padovani93}, Ho found $\dot{M}_{\star}\sim0.05 \, {\rm M}_{\odot}$ yr$^{-1}$. Therefore, stellar mass loss can sustain the gas reservoir in the NLR if the stellar debris survives for a lapse of time longer than $\sim 2\times10^{5}- 10^{6}$ yr before it dissipates and merges with the surrounding hot interstellar medium. The mass loss from evolved stars accounts for the high metallicity found in the ionized gas of Seyfert galaxies \citep{storchi98,hamann02}. Indeed, only in rare cases the metallicity appears to be sub-solar \citep{ghk06,izotov08}. 
Following \citet{ost06}, we determined the ionized gas mass by means of the Eq.\,\ref{equ:mass04}:
\begin{equation}
 M=6.67\times10^{-33}\, \frac{L_{\rm H\beta}}{N_{\rm e}} ~{\rm M}_\odot
\label{equ:mass04}
\end{equation}
By assuming the densities derived from the RS2 ratio, we obtained a mass distribution having a median value of $10^{6}$ M$_{\odot}$ and with 80 per cent of the sample having a mass lower than $3.4 \times10^{6}$ M$_{\odot}$. Deriving the NLR radius under the assumption of spherical distribution, we found a median value of 20 pc, with 80 per cent of the sample under $40$ pc. Of course these values are too low compared with the NLR radii. Even if a filling factor of $10^{-3}$ is assumed, the median radius is 200 pc. The 80 per cent of the sample has a radius lower than 400 pc, a more reasonable value but still rather low compared with estimated values typically found in the literature \citep[e.g.][]{bennert06b, krae09, yone06}. These results suggests that the NLR mass calculated with RS2 density values is probably underestimated. 
If we assume a low density medium (e.g. 1--10 cm$^{-3}$), the median value of the mass becomes about $2.5\times10^{7-8}$ M$_{\odot}$, and the median values of the radius changes between 200 and 900 pc, more in agreement with the observations. This simple exercise supports the idea that most of the NLR is made of a low density medium.
Alternatively, it is possible to determine the gas mass from the measure of the reddening and from a given dust-to-gas ratio \citep{fs07}. Assuming the Galactic dust-to-gas ratio we obtain the Eq.\,\ref{equ:mass05}:
\begin{equation}
 M=5.28\times 10^{21}\, {\rm cm}^{-2}\, {\rm mag}^{-1}\, {\rm E}(B-V)\, m_{\rm p}\, \Omega\, d^{2}_{\rm A}
\label{equ:mass05}\\
\end{equation}
where $\Omega$ is the solid angle subtended by the NLR and $d_{\rm A}$ the angular distance. 
Unfortunately it is not straightforward to apply this formula in our case.  
The fiber diameter is obviously fixed and it is not possible to get $\Omega$ from observational measurements. In principle one can assume fixed dimensions of the NLR, but this is not a good idea because the typical NLR radius varies in a wide range ($100$ to $1000$ pc or more). Nevertheless if we determine the NLR mass from Eq.\,\ref{equ:mass04}, Eq.\,\ref{equ:mass05} is useful to estimate the NLR dimensions. In fact by combining these equations we can express $\Omega\, d^{2}_{\rm A}$ as a function of measurable quantities. This gives us the projected NLR surface and then the NLR radius
\begin{equation}
 R=0.16~\sqrt{\frac {M}{{\rm A}(V)}} ~~{\rm pc}
\label{equ:radius}\\
\end{equation}
where tha mass $M$ is given in solar masses.
Table \ref{tab:NLRrad} contains the radius distributions, for both the samples, obtained with the three calculated mass median values through Eq.\,\ref{equ:radius}. In order to obtain the same radius distribution with both the approaches (Eq.\,\ref{equ:mass04} and \ref{equ:radius}) and with $10^{6}$ M$_{\odot}$ we have to impose a filling factor of  $\sim 0.001-0.01$ when the classical approach is used. 

As already pointed out by \citet{fs06}, and in agreement with what we said above, we can reasonably assume that most of the emitting volume of the NLR is dominated by a massive and highly ionized diffuse medium at very low density (1--10 cm$^{-3}$), containing less ionized clouds with density of $\sim$100--500 cm$^{-3}$, and likely clouds and/or filaments at higher density ($\sim 10^3$--10$^5$ cm$^{-3}$).

\begin{figure}
\centering
\includegraphics[width=84 mm]{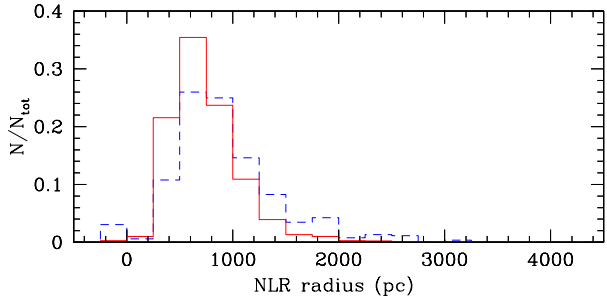}
\caption{\small NLR radii distributions using $N_{\rm e} = 10\, {\rm cm^{-3}}$. Colours and lines are like in Fig.\,\ref{fig:Av_cum}.}
\label{fig:NLRrad}
\end{figure}

\begin{figure}
 \centering
 \includegraphics[width=84 mm]{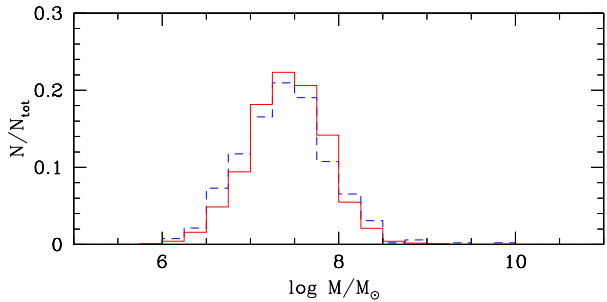}
 \caption{\small Ionized gas mass obtained by Eq.\,\ref{equ:mass04} with $N_{\rm e}=10\, {\rm cm^{-3}}$. Lines and colours are like in  Fig.\,\ref{fig:Av_cum}.}
 \label{fig:masses}
\end{figure}

\begin{table}
\caption{NLR mass (Eq.\,\ref{equ:mass04}) and radius (Eq.\,\ref{equ:radius}) distributions for three different $N_{\rm e}$.} 
\begin{center}
\tiny
\begin{tabular}{|l|l|l|l|l|}
\hline
 $N_{\rm e}$ cm$^{-3}$& M (M$_{\odot})$ & mean (pc)&  median (pc)& $<$80 per cent (pc) \\
\hline 
S2 & & & &\\
\hline
RS2 & $10^6$            & $170$  & $140$ & $240$ \\ 
$10$ & $2.5\cdot 10^{7}$  & $750$  & $680$ & $950$ \\ 
$1$ & $2.5\cdot 10^{8}$  & $2400$  & $2200$ & $3000$\\
\hline 
S-I & & & &\\
\hline
RS2 & $10^6$            & $200$  & $150$ & $260$ \\ 
$10$ & $2.5\cdot 10^{7}$  & $1000$  & $850$ & $1200$ \\ 
$1$ & $2.5\cdot 10^{8}$  & $3100$  & $2700$ & $4000$\\
\hline
\end{tabular}
\end{center}
\label{tab:NLRrad}
\end{table}
In order to compare the NLR radii and masses of the two samples, we assumed a density of 10 cm$^{-3}$. The NLR radii (Fig.\,\ref{fig:NLRrad}) are slightly larger in the S-I sample, likely because of lower A$(V)$ values measured in these objects (Fig.\,\ref{fig:Av_cum}). On the other hand, the mass distributions are similar for both the samples (Fig.\,\ref{fig:masses}).
The median value is about $3\times10^{7}$ M$_{\odot}$, and 90 per cent of the sample has a mass lower than $10^{8}$ M$_{\odot}$. 

\subsection{Kinematics}
The measured FWHM (FWHM$_{\rm m}$) was corrected for instrumental broadening to obtain the intrinsic value (FWHM$_{\rm i}$). As instrumental width we used the nominal SDSS spectral resolution, R = 1800 \citep{sdss_res} which corresponds to 167 km s$^{-1}$. 

 \begin{equation} 
FWHM_{\rm i}=\sqrt{(FWHM_{\rm m} \times {\rm c}/\lambda_{\rm i})^{2}-167^{2}}~~~{\rm km s}^{-1}
\end{equation}

\begin{figure}
 \centering
 \includegraphics[width=84 mm]{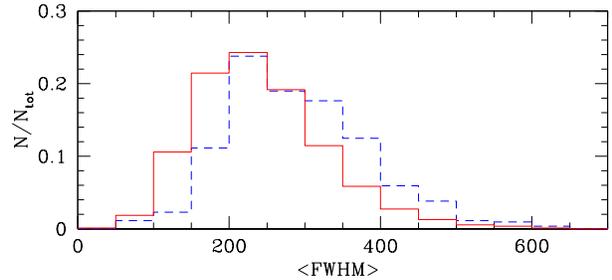}
 \caption{\small The distributions of the mean FWHM (km s$^{-1}$) of the emission-lines measured in each target. Lines and colours are like in Fig.\,\ref{fig:Av_cum}.}
 \label{fig:FWHM_tot_fam}
\end{figure}

\begin{figure}
\centering
 \includegraphics[width=75 mm]{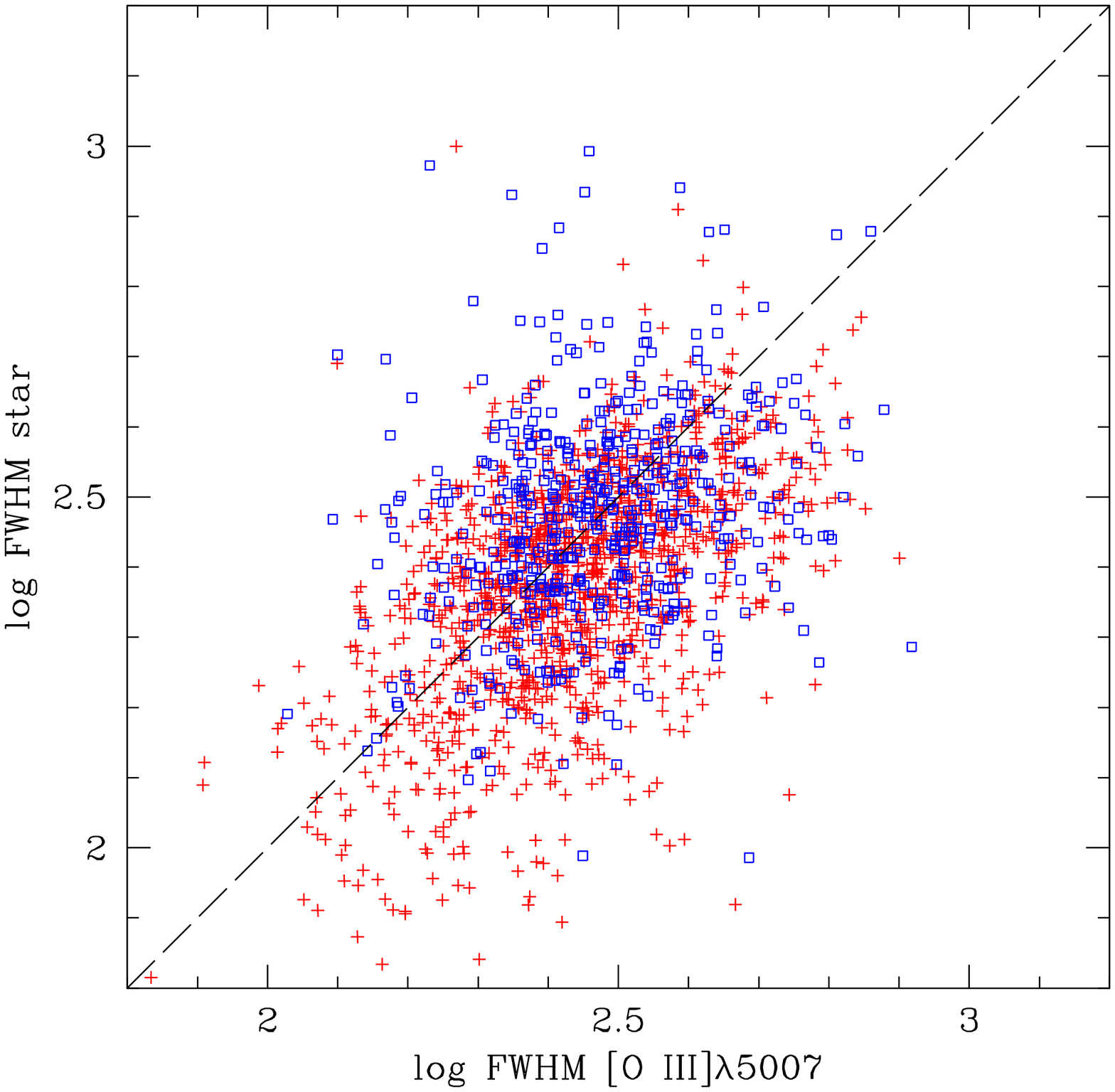}
 \includegraphics[width=75 mm]{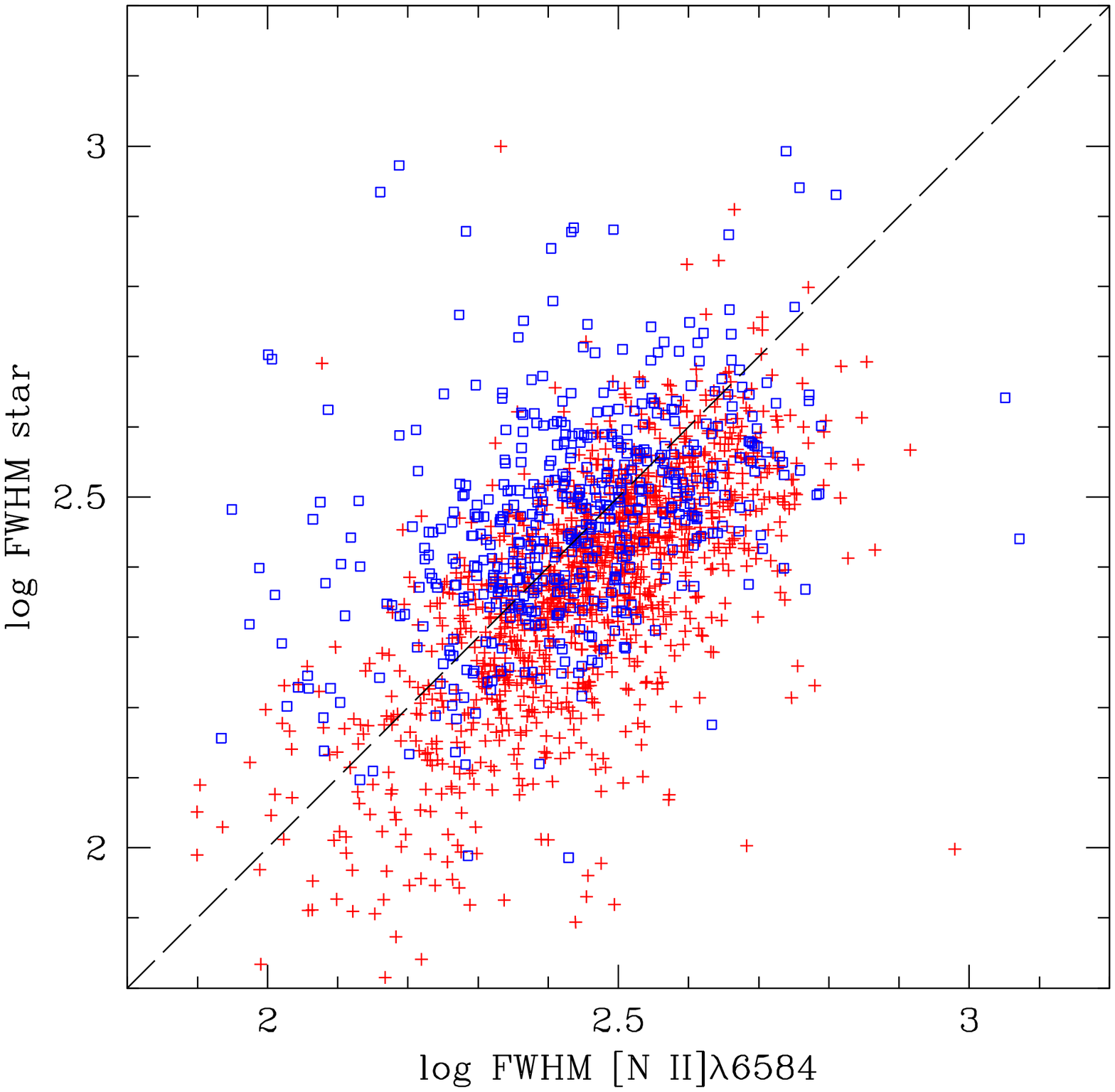}
 \caption{\small The FWHM of [O\,{\sc iii}]$\lambda$5007 (up) and [N\,{\sc ii}]$\lambda$6584 (bottom) are plotted against FWHM of the stellar component (FWHM$_{\rm star}$). 
Symbols are like in Fig.\,\ref{fig:O3HavsN2Ha}.}
 \label{fig:v_star_gas}
\end{figure}

Since a low S/N ratio makes the line profile difficult to be measured, we considered only the brightest lines, namely: H$\beta$, H$\alpha$, [O\,{\sc i}]$\lambda$6300, [N\,{\sc ii}]$\lambda$6584, [S\,{\sc ii}]$\lambda$6716,6731, [Ne\,{\sc iii}]$\lambda$3869, [O\,{\sc iii}]$\lambda$5007 and He\,{\sc ii} $\lambda$4686.
Furthermore, we calculated the mean FWHM for each object (see Fig.\,\ref{fig:FWHM_tot_fam}), using the previously mentioned lines. A very little difference between the samples is found by means of the KS test (D=0.08, $\Delta=0.15$, significance=1 per cent). The median values are about 270 km s$^{-1}$ and 300 km s$^{-1}$ respectively for S2 and S-I, with the 80 per cent of the samples showing FWHM lower than 360 and 390 km s$^{-1}$ respectively for S2 and S-I.
Concerning a possible connection between gas velocity dispersion and stellar velocity dispersion we obtained weak correlation with the exclusion of low ionization lines, which are formed, presumably, in the outer regions of the NLR. Then it is reasonable that at large distances from the nucleus the motion is governed by the gravitational field \citep{walsh08}. 
We compared the FWHM$_{\rm i}$ with the stellar velocity dispersion obtained by {\sc starlight}. 
The Spearman rank correlation coefficients between FWHM$_{\rm star}=2.35\sigma_{\star}$ and FWHM$_{\rm gas}$ are listed in Table \ref{tab:v_star_gas_bis}, with the usual meaning of the symbols. 
In S2 galaxies clear correlations exist except for [Ne\,{\sc iii}]$\lambda$3869 and He\,{\sc ii}$\lambda$4686, while in S-I galaxies only the Balmer lines show a correlation with stellar kinematics, even if weaker. We note that in the S2 sample, the low ionization lines show a better correlation, except for [O\,{\sc i}]$\lambda$6300 which is a rather weak line. 
In particular, we observed a stronger correlation for [N\,{\sc ii}]$\lambda$6584 (r$_s$=0.69) than for [O\,{\sc iii}]$\lambda$5007 (r$_s$=0.53). 
This confirms the results by \citet{nw96}, \citet{2005MNRAS.356..789B} and \citet{gh05}, that [O\,{\sc iii}]$\lambda$5007 is not a good indicator of stellar velocity dispersion. A better indicator is [N\,{\sc ii}]$\lambda$6584, as already shown by \citet{ho09}. 
The diagrams of FWHM$_{\rm star}$ vs. FWHM$_{\rm gas}$ are shown in Fig.\,\ref{fig:v_star_gas} for both [N\,{\sc ii}]$\lambda$6584 and [O\,{\sc iii}]$\lambda$5007. 

\begin{table}
\begin{center}
\caption{Correlation coefficients between the logarithms of FWHM$_{\rm star}$ and FWHM$_{\rm gas}$}
\begin{tabular}{lrrrrrr}
\hline
 &  & S 2 &    &  & S-I &  \\ 
 line & N & r$_s$ & $t$ &  N & r$_s$ & $t$ \\ 
\hline
 H$\beta$  & 2128 & 0.64   & 38.4    & 506 & 0.47 & 11.9\\      
 H$\alpha$ & 2126 & 0.71   & 46.5    & 511 & 0.53 & 14.1\\      
 6300      & 2097 & 0.56   & 30.9    & 508 & 0.33 & 7.8\\             
 6584      & 2124 & 0.69   & 43.9    & 509 & 0.47 & 12.0\\             
 6716      & 2106 & 0.66   & 40.3    & 506 & 0.36 & 8.7\\             
 6731      & 2103 & 0.62   & 36.2    & 504 & 0.32 & 7.6\\             
 4686      & 346  & 0.29   & 5.6     & 194 & 0.28 & 4.0\\           
 3869      & 1114 & 0.39   & 14.1    & 426 & 0.25 & 5.3\\           
 5007      & 2129 & 0.53   & 28.8    & 513 & 0.28 & 6.6\\             
\hline
\end{tabular}
\label{tab:v_star_gas_bis}
\end{center}
\end{table}

\begin{figure}
 \centering
 \includegraphics[width=84 mm]{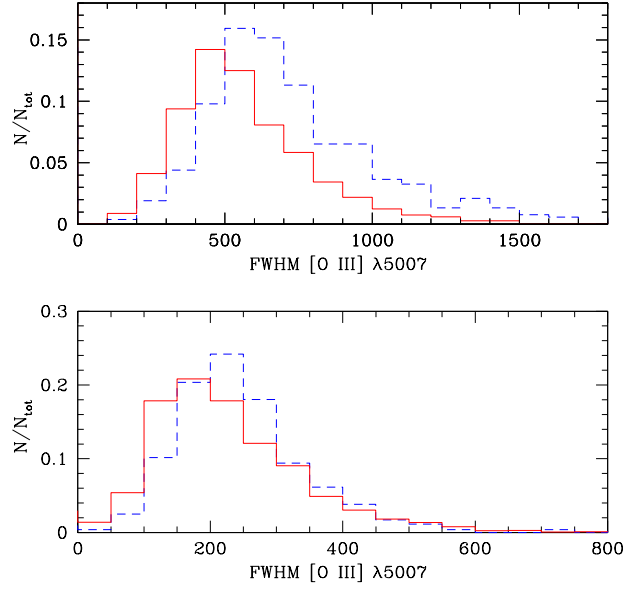}
 \caption{\small The distributions of the FWHM of [O\,{\sc iii}]$\lambda$5007 for the first component (bottom panel) and for the second component (top panel). Colours and lines are like in Fig.\,\ref{fig:Av_cum}.}
 \label{fig:FWHM_5007nb}
\end{figure}

\begin{figure}
 \centering
 \includegraphics[width=84 mm]{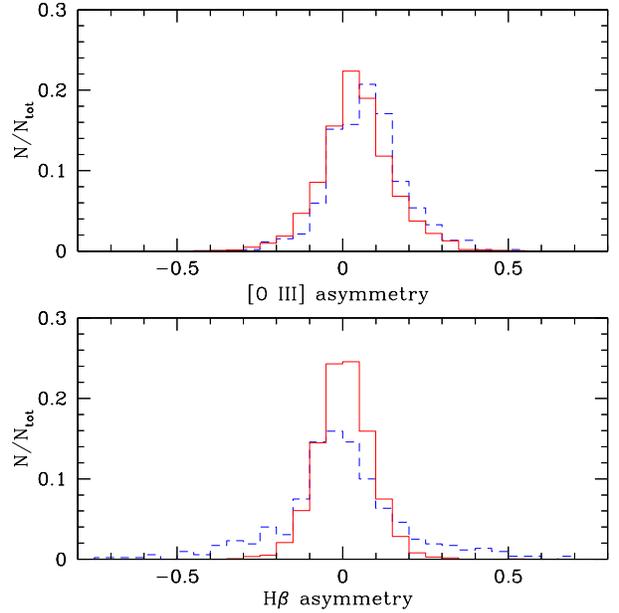}
 \caption{\small The distributions of the asymmetry for [O\,{\sc iii}]$\lambda$5007 (top panel) and for H$\beta$ (bottom panel). Colours and line styles are as in Fig.\,\ref{fig:Av_cum}.}
 \label{fig:asy}
\end{figure}

\begin{figure}
 \centering
 \includegraphics[width=84 mm]{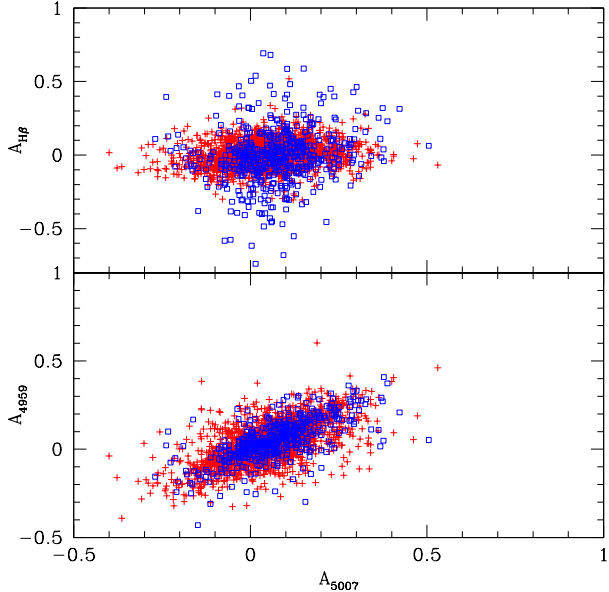}
 \caption{\small Asymmetry correlations, A$_{5007}$ vs. A$_{4959}$ (bottom panel), and A$_{5007}$ vs. A$_{\rm H\beta}$ (top panel). 
Symbols are like in Fig.\,\ref{fig:O3HavsN2Ha}.}
 \label{fig:asy_5007_4959_4861}
\end{figure}

In many cases, it was necessary to fit [O\,{\sc iii}]$\lambda$5007 with two components, a narrower (hereafter first component) and a broader component (hereafter second component). Fig.\,\ref{fig:FWHM_5007nb} shows the FWHM of [O\,{\sc iii}]$\lambda$5007 distributions for both the components. The values are systematically lower in the S2 sample with respect to the S-I sample. The largest difference is in the second component: the median value is 520 km s$^{-1}$ for S2 and 650 km s$^{-1}$ for the S-I sample. 80 per cent of the sample has a FWHM of the [O\,{\sc iii}] second component lower than 720 and 950 km s$^{-1}$ respectively for S2 and S-I.
An accurate analysis of H$\beta$ and [O\,{\sc iii}]$\lambda$4959,5007 profiles was performed. 
The asymmetry measure was obtained using the algorithm introduced by \citet{whi85}. Once the continuum level of the line is defined, the algorithm calculates the 10, 50 and 90 per cent of the area under the line profile and the corresponding wavelengths, from the blue side of the line. Then two line widths named respectively a=$\lambda_{50}-\lambda_{10}$ and b=$\lambda_{90}-\lambda_{50}$ are defined, so the asymmetry is calculated as A=(a-b)/(a+b). 
These measures are based on the area, so they are less sensitive to noise or effects of instrumental resolution. 
The main issue is the definition of the line profile limits to fix the continuum level. 
A number of different baselines was used and the values of the resulting parameters were averaged. In order to define the line boundaries and the continuum we applied the same method described in Section \ref{sec:spectra}. Once evaluated the line boundaries, we considered three pixels at each side of the line (the pixel corresponding to the line boundary and the adjacent ones) in order to define 9 different baselines. A maximum limit of $\pm$1500 km s$^{-1}$ was imposed to prevent the non-convergence: in these cases the continuum value plus 2 or 3 times $\sigma_{\rm c}$ was taken into consideration, especially for [O\,{\sc iii}]4959 and H$\beta$, which have sometimes a relatively low S/N ratio. 
The distributions of the measured asymmetries are presented in Fig.\,\ref{fig:asy}. A clear indication of a positive asymmetry (blue wing) of the [O\,{\sc iii}]$\lambda$5007 line with respect to H$\beta$ is found.  
Of course the H$\beta$ asymmetry distribution for the S-I sample is totally different because of the broad component. However, the mean value for both samples is around A=0. In order to check the results, the asymmetry analysis was performed also for [O\,{\sc iii}]4959. There is agreement between [O\,{\sc iii}]4959 and [O\,{\sc iii}]$\lambda$5007 (Fig.\,\ref{fig:asy_5007_4959_4861}) although, the asymmetry measure is more difficult in case of [O\,{\sc iii}]$\lambda$4959, because this line is weaker. The Spearman rank correlation coefficients are r$_s$=0.60 ($t=34.6$, N=2131) for the S2 and r$_s$=0.68 ($t=21.0$, N=514) for the S-I sample. On the other hand there is no correlation between [O\,{\sc iii}]$\lambda$5007 and H$\beta$. 
The [O\,{\sc iii}]$\lambda$5007 asymmetry parameter gives highly positive asymmetry for 68 per cent of the S2 and about 74 per cent of the S-I.
The KS test gives a very little difference between the samples (D=0.08, $\Delta=0.14$, significance=1 per cent).
This result supports the idea that when the asymmetries are present we are looking in the inner regions of the NLR where the kinematics is more complicated and chaotic, and probably the gas is not simply moving in the gravitational potential but it is driven by winds and decelerating outflows \citep[]{wagner97,kom08}.
Finally, we did not find any correlation between these asymmetry parameters and the luminosity, the FWHM of both components, and A$(V)$.

\subsection{Energy and Accretion rate}
\begin{figure}
 \centering
 \includegraphics[width=84 mm]{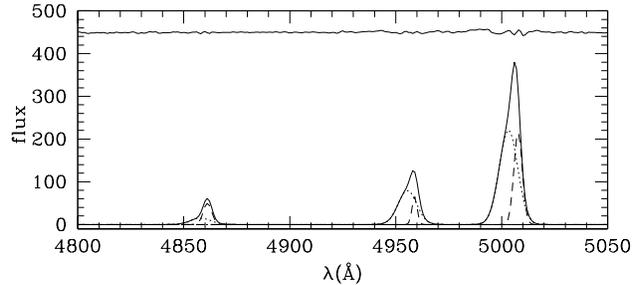}
 \caption{\small An example of double components fitting: the observed spectrum (solid line), the narrow component (dashed line) and the broader component (dotted line). Residuals are plotted on the top of the diagram.}
 \label{fig:2ndcomp}
\end{figure}

With the ionized gas mass and velocity values, it is possible to determine the kinetic energy due to turbulence, thermal and bulk motion.
The third one would require velocity field, which are not available in our case. We could only use the asymmetric components (broader or second component) found in the [O\,{\sc iii}]$\lambda$5007 line to derive the order of magnitude of the kinetic energy due to bulk motion. These energies are determined using the formalism introduced by \citet{fs07}.
The turbulent energy can be estimated from the measured velocity dispersion $\sigma$(H$\beta)$,
\begin{equation}
 E_{\rm tur}=M \sigma^{2}=5.0\times10^{56} M_{10}\, \sigma^{2}_{50}~~ {\rm erg}
\label{equ:Etur}
\end{equation}
where $\sigma_{50}$ = $\sigma ({\rm H}\beta)/50$ km s$^{-1}$ and M$_{10}$ the mass expressed in $10^{10}$ M$_{\odot}$.
The thermal energy is 
\begin{equation}
 E_{\rm th}=\frac{3}{2} M{\rm k}T / m_{\rm p}= 2.5\cdot 10^{55} M_{10}\,T_{4} ~~{\rm erg}
\label{equ:Eth}
\end{equation}
where $T_{4} = T/10000K$
The bulk kinetic energy is given by
\begin{equation}
 E_{\rm bulk}=\frac{1}{2} MV^{2}= 6.2\times 10^{57} M_{10}\,V^{2}_{250} ~~{\rm erg}
\label{equ:bulk}
\end{equation}
where $V_{250}$ is the bulk velocity in $V/250$ km s$^{-1}$ units. The bulk velocity is estimated from the difference in the peak velocities of the [O\,{\sc iii}]$\lambda$5007 components ($\delta$p parameter), the narrow component is assumed to have the systemic velocity (Fig.\,\ref{fig:2ndcomp}).

\begin{figure}
 \centering
 \includegraphics[width=84 mm]{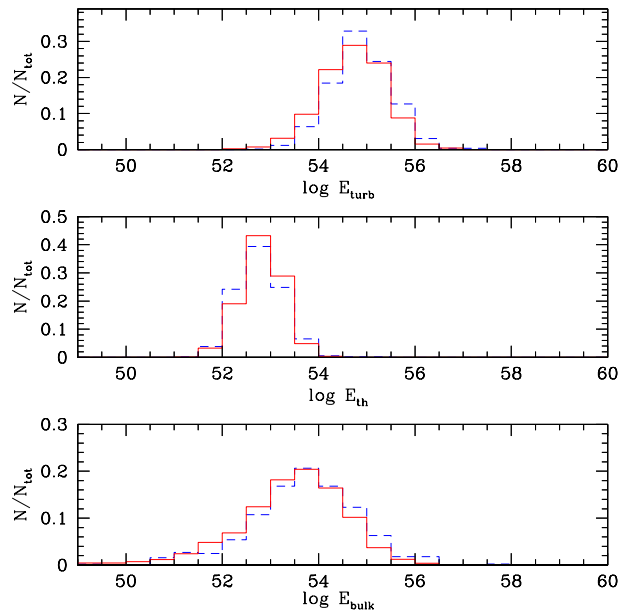}
 \caption{\small The distributions of turbulent energy (top panel), thermal energy (middle panel) and bulk energy (bottom panel) in erg}. Colours and lines are as in Fig.\,\ref{fig:Av_cum}.
 \label{fig:energy_dist}
\end{figure}

\begin{figure}
\centering
\includegraphics[width=84 mm]{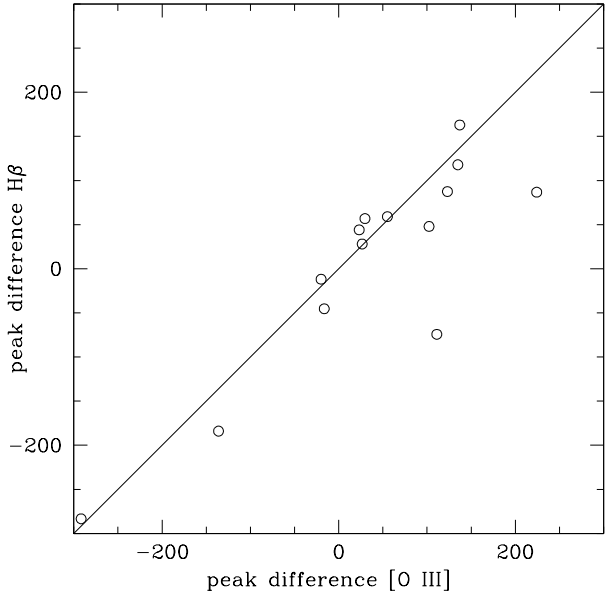}
\caption{\small The peak difference velocities of [O\,{\sc iii}]$\lambda$5007 (km s$^{-1}$) are plotted against the same for H$\beta$, for spectra with 2 components in both the lines (see text).}
\label{fig:2nd_ratio}
\end{figure}

\begin{figure}
\centering
\includegraphics[width=84 mm]{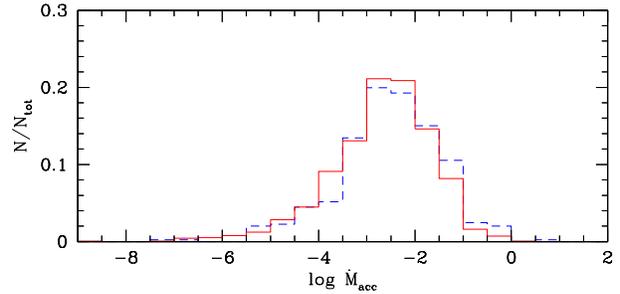}
\caption{\small Mass accretion rate distributions. Colours and lines are like in Fig.\,\ref{fig:Av_cum}.}
\label{fig:Macc}
\end{figure}

Thermal and turbulent energy distributions are plotted in Fig.\,\ref{fig:energy_dist}. 
The distributions are similar for S2 and S-I with a logarithmic median value around 52.7 and 54.7 respectively for the thermal and turbulent energy. 
The thermal energy is negligible compared to the turbulent energy.
In order to derive the bulk kinetic energy it is necessary to obtain the [O\,{\sc iii}]$\lambda$5007/H$\beta$ ratio of the second component and then apply Eq.\,\ref{equ:mass04} to estimate the involved mass. Unfortunately, only in 14 spectra it was possible to fit two components in both [O\,{\sc iii}] and H$\beta$. However, by plotting $\delta$p (H$\beta$) vs. $\delta$p ([O\,{\sc iii}]), we found a good agreement between the fitted values, and the flux ratio of the second component is between 3 and 10 (Fig.\,\ref{fig:2nd_ratio}), which is consistent with the assumption of \citet{fs06} and with the measures of \citet{cid01}. 
Then, by assuming 10 as a representative value of the [O\,{\sc iii}]$\lambda$5007/H$\beta$ ratio, from the luminosity of [O\,{\sc iii}]$\lambda$5007 second component and assuming low density $N_{\rm e}=10$ cm$^{-3}$ we obtained a mass distribution with a median logarithmic value around 6.8 for both the samples.
Hence the masses involved in the bulk motions are about 1/3 of the total mass. If the bulk motions are connected to outflows and we assume a typical dynamical time scale of $10^{7}$ yr (see below), the resulting rate of mass outflow is $\sim1$ M$_{\odot}$ yr$^{-1}$.
Finally, from the velocity peak difference, we derived the bulk kinetic energy by means of Eq.\,\ref{equ:bulk}. 
The logarithm of median bulk kinetic energy values is around 53.5. The distributions (Fig.\,\ref{fig:energy_dist}, bottom panel) show that the 80 per cent of the samples have values lower than 54.5, this means that the bulk kinetic energy is approximately lower than one order of magnitude compared to the turbulent energy. 
We stress that this could be an upper limit, since we assumed that all the flux of the second component is emitted by outflowing mass.
If we assume that the outflow is directly driven by the black hole (BH), we can estimate the mass accretion rate necessary to sustain the outflow.
The input momentum rate from radiation pressure is:
\begin{equation}
 \dot{p}=1.3 \times 10^{35} (\eta/0.1) \dot{M}_{\rm acc} ~~~{\rm dyne}
\label{equ:mom_rate}
\end{equation}
where $\eta$ is the radiative efficiency and $\dot{M}_{\rm acc}$ is the mass accretion rate of the BH in units of M$_{\odot}$, yr$^{-1}$ \citep{fs06}. 
The momentum rate of the moving bulk mass is given by:
\begin{equation}
 \dot{p} \sim M\times V_{\rm bulk}/t_{\rm dyn}~~~{\rm dyne}
\label{equ:mom_rate_bis}
\end{equation}
where $t_{\rm dyn}$ is the dynamical time scale. 
We estimated $t_{\rm dyn}$ from the NLR radius
\begin{equation}
t_{\rm dyn}\sim R_{\rm NLR}/V_{\rm bulk} ~~~{\rm s}
\label{equ:tdyn}
\end{equation}
obtaining the same distributions for both samples.
The median value is $10^{7}$ yr, and 80 per cent of the samples have $t_{\rm dyn}<3\times 10^{7}$ yr. 
Finally, by assuming $\eta\sim 0.1$ and making Eq.\,\ref{equ:mom_rate} equal to Eq.\,\ref{equ:mom_rate_bis}, we obtained an estimate of the $\dot{M}_{\rm acc}$ distribution (Fig.\,\ref{fig:Macc}). The median value is around 0.003 M$_{\odot}$ yr$^{-1}$, 80 per cent of the samples have $\dot{M}_{\rm acc}<0.03$ M$_{\odot}$ yr$^{-1}$. 
These values are in agreement with the generally accepted accretion rates in Seyfert galaxies.

\section{Summary and Conclusions}
A sample of 5678 Seyfert galaxies was selected from SDSS-DR7 by applying the O$_{123}$ diagram and their emission-lines were measured by means of a dedicated automatic code based on a nonlinear least square fitting method. Two components were applied to H$\beta$ and H$\alpha$ when necessary, and to [O\,{\sc iii}]4959,5007 emission-lines in order to analyze their profiles. The galaxies were divided into two groups on the basis of the H$\alpha$ FWHM values and of the VO diagnostic diagrams, obtaining 2153 Seyfert 2 galaxies and 521 Intermediate-type Seyfert galaxies. 
Our results, obtained through a detailed spectroscopic analysis of the NLR of these two samples and a comparison between their physical properties, are in agreement with what is expected by the Unified Model and add new details to the general knowledge of the AGN properties.

\begin{enumerate}
 \item The reddening for the gas, obtained by means of the Balmer decrement, is on average stronger in S2 ($<$A($V$)$>$ $\sim$ 1.5) than in S-I ($<$A($V$)$>$ $\sim$ 1).
According to the Unified Model this is expected, because of the different line-of-sight orientation with respect to the torus axis. 
Indeed, \citet{2000ApJS..126...63R} found ${\rm A}(V)$ = 0.663 $\pm$ 0.345 for a sample of 16 Broad Lined Seyfert 1 (BLS1).
In addition, the reddening of the stellar component shows lower A$(V)$ values, weakly correlated with those for the gas. This suggests that the extinction is mostly caused by dust associated to the gas of the NLR.

\item Both the samples show good correlation between the lines having similar ionization potential.

\item The absence of [Ar\,{\sc iv}]$\lambda$4711,4740 doublet and the values of temperature derived by [O\,{\sc iii}], [O\,{\sc ii}] and [S\,{\sc ii}] line ratios, strongly support the idea that the NLR of Seyfert galaxies is a multi-phase ionized gas. In particular, it is likely made of a very low density medium ($N_{\rm e} \sim 1-10$ cm$^{-3}$), where filaments or clouds with low density ($N_{\rm e} \sim 10^2$ cm$^{-3}$) and high density ($N_{\rm e} > 4\times10^5$ cm$^{-3}$) are moving. The high density is expected to suppress [Ar\,{\sc iv}] lines and at the same time to pump [O\,{\sc iii}]$\lambda$4363 by collisional processes.

\item The [Fe\,{\sc vii}] coronal lines are detected in $\sim 4$ per cent of the S2 sample, and $\sim 19$ per cent of the S-I sample. 
The analysis of the [Fe\,{\sc vii}]$\lambda$6087/[O\,{\sc iii}]$\lambda$5007 emission-line ratio indicates higher values for the S-I sample and suggests that the coronal lines are formed deep inside the NLR.
This result is in agreement with \citet{2000AJ....119.2605N} who found $\log$[Fe\,{\sc vii}]$\lambda$6087/[O\,{\sc iii}]$\lambda$5007 = -0.94$\pm$0.39, -1.33$\pm$0.43, -1.81$\pm$0.29 for BLS1, S-I and S2, respectively.

\item The ionization parameter $U$ is in general significantly higher in S-I than in S2, showing average values of about $-2.4$ and $-2.7$, respectively. However, if we isolate the S2 galaxies showing high ionization lines, as He\,{\sc ii} $\lambda$4686 and/or [Fe\,{\sc vii}]$\lambda$6087, the distributions of $U$ for the two samples become similar. This result can be explained in case of a thinner torus which let the deeper and more strongly ionized NLR to be observed.
A comparison can be made with \citet{1983ApJ...264..105F} who estimated typical values of -2, -2.5 and -3.5, for S1, S2 and LINERs, respectively, by means of photoionization models.

\item A good correlation between gaseous and stellar kinematics is observed only in S2 galaxies, and only in case of low ionization and Balmer lines. This suggests that the kinematics of the NLR is dominated by the gravitational potential of the galaxy only in the less ionized regions. Therefore, we confirm and point out that the FWHM of [O\,{\sc iii}]$\lambda$5007 line cannot be use as a surrogate of the stellar velocity dispersion in Seyfert galaxies. 

\item In addition, about 75 per cent of galaxies in both samples show asymmetries on the blue wing of the [O\,{\sc iii}]$\lambda$5007 line profile, suggesting the presence of radial motions, inflows and/or outflows, of the highly ionized gas. Also H$\beta$ lines show asymmetries, which are nevertheless not correlated with those observed in [O\,{\sc iii}], with the exception of few cases ($<10$ per cent).
[O\,{\sc iii}] $\lambda$5007 profiles with strong blue wings have been also found in a sample of radio-loud quasars by \citet{1996ApJS..102....1B}. The author claimed that the H$\beta$ profiles show strong red wings. 

\item However, energy balance calculations, based on the estimated ionized gas mass, which is on average around $3\times10^7$ M$_{\odot}$ for both samples, clearly indicate that the turbulent energy is the dominant component, with values in the range $10^{54} - 10^{56}$ erg. The kinetic bulk energy is an order of magnitude lower, while the thermal energy plays a negligible role.
\end{enumerate}

\section*{Acknowledgments}
We are greatful to the anonymous referee for useful comments and suggestions which improved the quality of the paper.

This research has made use of the NASA/IPAC Extragalactic Database (NED) which is operated by the Jet Propulsion Laboratory, California Institute of Technology, under contract with the National Aeronautics and Space Administration. 

Funding for the SDSS and SDSS-II has been provided by the Alfred
P. Sloan Foundation, the Participating Institutions, the National
Science Foundation, the US Department of Energy, the National
Aeronautics and Space Administration, the Japanese Monbukagakusho,
the Max Planck Society and the Higher Education Funding
Council for England. The SDSS Web Site is http://www.sdss.org/.
The SDSS is managed by the Astrophysical Research Consortium
for the Participating Institutions. The Participating Institutions are
the American Museum of Natural History, Astrophysical Institute
Potsdam, University of Basel, Cambridge University, Case Western
Reserve University, University of Chicago, Drexel University,
Fermilab, the Institute for Advanced Study, the Japan Participation
Group, Johns Hopkins University, the Joint Institute for Nuclear Astrophysics,
the Kavli Institute for Particle Astrophysics and Cosmology,
the Korean Scientist Group, the Chinese Academy of Sciences
(LAMOST), Los Alamos National Laboratory, the Max-Planck-
Institute for Astronomy (MPA), the Max-Planck-Institute for Astrophysics
(MPIA), New Mexico State University, Ohio State University,
University of Pittsburgh, University of Portsmouth, Princeton
University, the United States Naval Observatory and the University
of Washington.

\label{lastpage}

\end{document}